\documentclass[aps,prb,twocolumn,amsmath,amssymb,superscriptaddress,showpacs]{revtex4}
\usepackage{xcolor}
\usepackage{textcomp}
\usepackage{graphicx}
\usepackage{esint}
\PassOptionsToPackage{normalem}{ulem}
\usepackage{ulem}
\usepackage[unicode=true,pdfusetitle,
 bookmarks=true,bookmarksnumbered=false,bookmarksopen=false,
 breaklinks=false,pdfborder={0 0 0},backref=false,colorlinks=true]
 {hyperref}
\hypersetup{colorlinks=true, citecolor=blue, urlcolor=blue, linkcolor=blue}

\begin{document}

\title{Adsorption geometry and electronic properties of flat-lying monolayers
of tetracene on the Ag(111) surface}

\author{N. L. Zaitsev}
\email{nza@yandex.ru}
\affiliation{Philipps-Universität Marburg, D-35032, Marburg, Germany}

\author{I. A. Nechaev}
\affiliation{Centro de F\'{i}sica de Materiales CFM-MPC, Centro Mixto CSIC-UPV/EHU,
20018, San Sebasti\'{a}n, Spain}
\affiliation{Tomsk State University, 634050, Tomsk, Russia}
\affiliation{Saint Petersburg State University, 198504, Saint Petersburg, Russia}

\author{U. H\"{o}fer}
\affiliation{Philipps-Universität Marburg, D-35032, Marburg, Germany}
\affiliation{Donostia International Physics Center (DIPC), 20018, San Sebasti\'{a}n,
Spain}

\author{E. V. Chulkov}
\affiliation{Donostia International Physics Center (DIPC), 20018, San Sebasti\'{a}n,
Spain}
\affiliation{Departamento de F\'{i}sica de Materiales, Facultat de Ciencias Qu\'{i}micas,
UPV/EHU Apdo. 1072, 20080, Donostia/ San Sebasti\'{a}n, Spain}
\affiliation{Centro de F\'{i}sica de Materiales CFM-MPC, Centro Mixto CSIC-UPV/EHU,
20018, San Sebasti\'{a}n, Spain}
\affiliation{Tomsk State University, 634050, Tomsk, Russia}
\affiliation{Saint Petersburg State University, 198504, Saint Petersburg, Russia}

\begin{abstract}
The geometrical and electronic properties of the monolayer (ML) of
tetracene (Tc) molecules on Ag(111) are systematically investigated
by means of DFT calculations with the use of localized basis set.
The bridge and hollow adsorption positions of the molecule in the
commensurate $\gamma$-Tc/Ag(111) are revealed to be the most stable
and equally favorable irrespective to the approximation chosen for
the exchange-correlation functional. The binding energy is entirely
determined by the long-range dispersive interaction. The former lowest
unoccupied orbital remains being unoccupied in the case of $\gamma$-Tc/Ag(111)
as well as in the $\alpha$-phase with increased coverage. The unit
cell of the $\alpha$-phase with point-on-line registry was adapted
for calculations based on the available experimental data and the
computed structures of the $\gamma$-phase. The calculated position
of the Tc/Ag(111) interface state is found to be noticeably dependent
on the lattice constant of the substrate, however its energy shift
with respect to the Shockley surface state of the unperturbed clean
side of the slab is sensitive only to the adsorption distance and
in good agreement with the experimentally measured energy shift.
\end{abstract}

\pacs{
68.43.-h,   
73.20.-r,   
73.20.At   
}

\maketitle

\section{introduction}

Organic molecular thin films are currently of great interest because
of their possible applications in micro- and optoelectronic devices
\cite{dimitrakopoulos_organic_2002,hoppe_organic_2004}. Their properties
depend on the nature of the interface between the molecular layer
and the substrate \cite{shen_how_2004,koch_moleculemetal_2013}. The
performance of the molecular devices is considerably conditioned by
the efficiency of charge transfer across the interface, which in turn
is governed by the relative alignment of molecular energy levels with
respect to the Fermi level of the metal substrate as well as the overlap
between molecular and substrate wave functions \cite{koch_energy_2008,tautz_structure_2007}.
The presence of interface electronic states (ISs) \cite{temirov_free-electron-like_2006,schwalb_electron_2008,zaitsev_change_2010}
is an additional agent influencing the overall charge transfer, albeit
role of these states in the process and the mechanism of their formation
are not yet fully understood \cite{marks_formation_2014}. On one
hand, such hybrid states can be formed as the result of the chemical
interaction of molecular orbitals with metallic states \cite{temirov_free-electron-like_2006,ferretti_mixing_2007,gonzalez-lakunza_formation_2008,ziroff_hybridization_2010}.
On the other hand, the symmetry breaking at the metal/organic interface
alone, can, in many cases, lead to new interface electronic states,
analogues to the Shockley state of clean metal surfaces \cite{schwalb_electron_2008,marks_energy_2011,galbraith_formation_2014,caplins_metal/phthalocyanine_2014,schafer_lifetimes_2000}.
Theoretical investigations of these types of organic/metal interface
states (IS) have focused on molecules with a perylene core and carboxylic
end groups, i.e. NTCDA and PTCDA \cite{zaitsev_change_2010,dyer_nature_2010,marks_energy_2011,zaitsev_transformation_2012,galbraith_formation_2014}.
The calculations reveal that the IS has its maximum probability density
between the top-most metallic layer and the plane of carbon atoms.
The IS wave function shows a similar penetration into the metal substrate
as the Shockley surface state \cite{chulkov_quasiparticle_2001}.
At the same time, the lateral corrugation of the IS local density
of states above the metal substrate resembles that of organic molecular
orbitals \cite{zaitsev_change_2010,dyer_nature_2010,marks_energy_2011,zaitsev_transformation_2012,galbraith_formation_2014,caplins_metal/phthalocyanine_2014}.

The interaction of NTCDA and PTCDA with many metal substrates, however,
is not of purely van-der-Waals type. On Ag(111) and Ag(100) a lowering
and partial filling of the lowest unoccupied molecular orbital of
the molecules is observed \cite{willenbockel_interplay_2015}. In
such a situation, it is difficult to assess to what extent this type
of chemical hybridization influences the interface state and whether
it attains similar properties for the case of physisorption. Here,
we thus consider a model system with a weaker molecule-substrate interaction,
Tetracene/Ag(111).

Tetracene (Tc) is an organic molecule with planar aromatic structure
($C_{18}H_{12}$ ) and it is one of the most promising organic semiconductor
for the application due to its high charge carrier mobility \cite{boer_field-effect_2003,newman_transport_2004}.
A number of the arrangement patterns (or ordered phases) of Tc molecules
on the Ag(111) surface has been observed as a function of coverage
\cite{soubatch_structure_2011,sueyoshi_spontaneous_2013}. The compressed
monolayer (ML) $\alpha$-phase with point-on-line type of commensurability,
which can undergo spontaneous structural transformation\cite{sueyoshi_spontaneous_2013},
has been studied in detail by different experimental techniques \cite{langner_structural_2005,gonella_tetracene_2008}.
Additionally, the fully commensurate $\gamma$-phase with submonolayer
coverage has been discovered rather recently \cite{soubatch_structure_2011}.
Both phases are characterized by the same orientation of the molecule
with respect to the substrate and they are suitable for the computational
study of the coverage impact on the adsorption geometry and electronic
properties. This work is intended to get a better understanding about
the interaction between Tc molecules and the silver (111) surface,
as well as its effect on the energy levels alignment and the interface
electronic states.

In this study, we have examined $\gamma$- and $\alpha$-Tc/Ag(111)
by means of density functional theory (DFT) calculations with numerical
pseudoatomic orbitals. Because of the large size of the system the
localized basis functions offer obvious advantage over the plane waves,
especially for the reasonable description of the interface state which
requires the metal surface to be represented by quite thick slab\cite{zaitsev_transformation_2012}.
Nevertheless, accurate description of the delocalized surface states
needs careful handling with pseudoatomic orbitals \cite{garcia-gil_optimal_2009}.
Here we explore a few ways of representation of the substrate wave
function by localized numerical basis functions and show that the
better one is to use different cutoff radii of the basis functions
for bulk and surface silver atoms.

It was revealed that the surface state energy of the clean surface
directly depend on its lattice constant. The same trend is demonstrated
by the interface state which appeared higher in energy of the hybrid
Tc/Ag(111) system, namely its absolute position increases with expansion
of the substrate. The binding between Tc and Ag(111) can be described
correctly only with taking into account the long-range dispersive
forces fully determining the substrate-adsorbate interaction in this
interface. In general, our calculations provide good description of
the available experimental data.

\section{methods}

The first-principles electronic structure calculation is performed
within the DFT as implemented in the SIESTA code \cite{ordejon_self-consistent_1996,soler_siesta_2002}.
Localized pseudoatomic orbitals were used for the wave function representation,
and deep core potentials were represented by norm-conserving Troullier-Martins
pseudopotentials \cite{troullier_efficient_1991} in the Kleynman-Bylander
\cite{kleinman_pseudopotentials_1982} nonlocal form. The conventional
generalized gradient approximation (GGA) was chosen for the exchange-correlation
functional with the PBE \cite{perdew_generalized-gradient_1996} parametrization.
Long-range dispersion forces were introduced by using the optB88-vdW
functional of the vdW-DF2 approach especially adapted for solids \cite{klimes_chemical_2010,klimes_van_2011}. 

The double-$\zeta$ polarized (DZP) basis set with the energy shift
of 10 meV and generated by a soft confinement scheme was used for
hydrogen (with the cutoff radius $r_{c}$ of 7.2~a.u.), oxygen (5.95~a.u.)
and silver (9.73~a.u.) atoms. Such basis functions for silver provide
the enlarged equilibrium lattice constant $a=4.23$~Å for both the
GGA-PBE and the optB88-vdW functional. The indicated cutoff-radius
value for silver was taken to reach the convergence with respect to
the lattice constant. However, the reasonable silver bulk properties
were found with the shorter radius $r_{c}=7.03$~a.u. of the basis
orbitals. Actually, the equilibrium lattice constant $a=4.16$~Å
and the bulk modulus $B=114$~GPa were obtained with the GGA-PBE
functional, and similar values ($a=4.17$~Å~, $B=119$~GPa) with
the optB88-vdW one. 

On the one hand, for a good description of the silver (111) surface,
i.e., to obtain accurate values of the surface energy, work function,
energy of the surface state and its decay into the vacuum, one needs
to use the basis functions with a large $r_{c}$ \cite{garcia-gil_optimal_2009}.
It was also shown \cite{lee_ab_2005} that the long range orbitals
are needed to minimize basis set superposition error (BSSE) and reproduce
the binding energy of plane wave calculations. On the other hand,
a shorter $r_{c}$ provides the bulk properties consistent with the
experiment, which are important to be described properly, because
the increasing of the lattice constant causes the up-shift in energy
of the surface state position of the Ag(111) surface. Moreover, as
calculations show, the adsorption distance of aromatic molecules on
metallic surfaces depends on the parameter $r_{c}$ of the substrate:
the shorter the radius we chose the smaller the distance we have \cite{buimaga-iarinca_adsorption_2014}.
Thereby, in examining the geometrical and electronic properties of
the metal-organic interface under study, we consider three approaches
to the silver-substrate description.

\begin{figure}
\includegraphics[width=1\columnwidth]{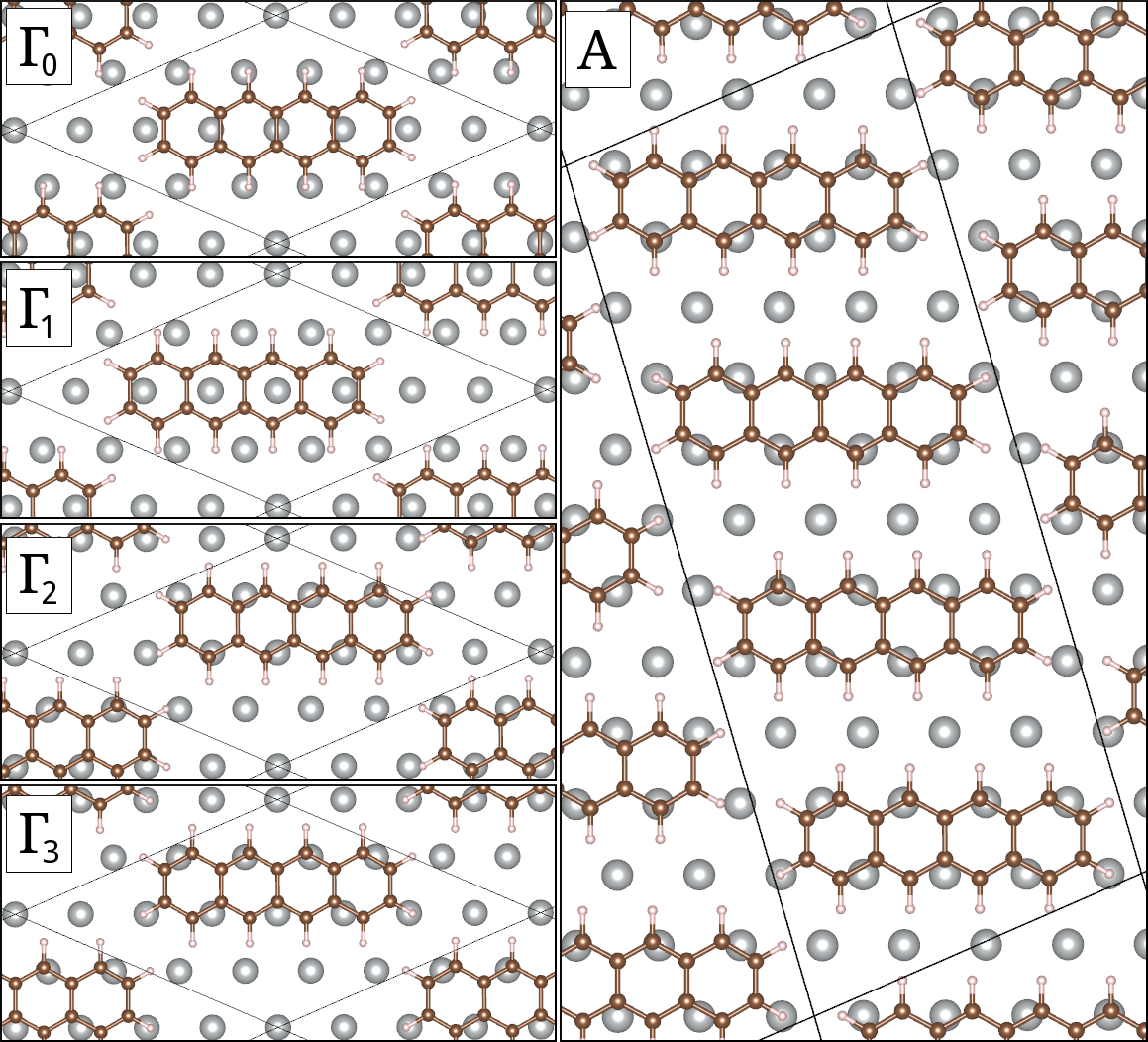}

\caption{\label{fig:tc-geo} Different adsorption
sites of $\gamma$-Tc/Ag(111). $\Gamma_{0}$ --- top site, $\Gamma_{1}$
--- long bridge site, $\Gamma_{2}$ --- bridge site and $\Gamma_{3}$
--- hollow site. The commensurate unit cell with four Tc molecules
represents the monolayer of $\alpha$-Tc/Ag(111).
}
\end{figure}

\emph{In the first one}, the experimental value for the silver lattice
constant ($a=4.09$~Å) and the largest radius of the basis functions
($r_{c}$=9.73 a.u.) are used. All silver atoms in the slab are pinned
to their bulk positions, but the atoms of the molecular monolayer
are allowed to be relaxed. \emph{In the second approach,} two types
of basis functions are used for silver atoms: $r_{c}=9.73$~a.u.
for atoms in the upper- and lowermost silver layers of the slab (the
external atoms) and $r_{c}=7.03$~a.u. for the rest of atoms (the
internal atoms). As was shown in Ref. \cite{garcia-gil_optimal_2009},
such an approach improves the description of the clean surfaces. The
lattice constant is set to its equilibrium value as found with the
basis functions of the internal atoms. Along with atoms in the molecular
ML, the positions of the atoms in the two uppermost silver layers
are optimized as well. We improve thus the description of surface
electronic bands and substrate bulk properties simultaneously. It
is worth noting that the use of orbitals with the largest $r_{c}$
for all atoms in the slab leads to insignificant changes of the electronic
band structure, but vastly increases the computational cost. In order
to demonstrate it, in\emph{ the third }type of substrate\emph{ }handling
all silver atoms are taken with the large $r_{c}=9.73$~a.u. In this
case, the equilibrium lattice constant has the value of $a=4.23$~Å.
Again, only the positions of the silver atoms in the two uppermost
layers of the slab are allowed to be optimized. Hereinafter, we refer
to these approaches of the substrate description as to Model 1, 2,
and 3, respectively.

The scheme of periodically repeated slabs are used to describe the
infinite close packed face-centered cubic Ag(111) surface. Interaction
between the periodic images of the systems in the direction perpendicular
to the surface (in the $z$-direction) is suppressed by the large
size of the cell along this direction, imposing a vacuum layer of
about 11~Å. A uniform mesh for the numerical integration and solution
of the Poisson equation is specified by the energy cutoff of 250 Ry.
The substrate was represented by four layers during structural relaxation
and by 12 layers for calculations of the interface band structure.
The molecular monolayer is applied to one side of the substrate only.
We use the k-point sampling of the surface Brillouin zone based on
the Monkhorst-Pack scheme with 22 k-points. The iterative modified
Broyden procedure \cite{johnson_modified_1988} is applied to reach
stable structures. All the considered geometries are relaxed until
all interatomic forces were smaller than 0.02 eV/Å. 

The spatial distribution of the interface state was computed with
the help of the OpenMX (version 3.7) DFT code \cite{_openmx_37,ozaki_variationally_2003,ozaki_numerical_2004,ozaki_efficient_2005}.
For silver atoms two types of basis functions with different cutoff
radii of 7.0 and 9.0 Bohr (as in Model 2), but with the same size
s2p2d2f1 were used, while for hydrogen and carbon atoms we chose H7.0-s2p1
and C7.0-s2p2d1, respectively. The notation in the last case means
that two primitive orbitals for each \emph{s} and \emph{p} states
and one primitive orbital for the \emph{d} states with the cutoff
radius of 7.0 Bohr were used. The real-space grid for numerical integration
was specified by the energy cutoff of 250 Ry. The total-energy convergence
was better than 0.027 meV. The surface Brillouin zone (SBZ) of the
supercell was sampled with a 6 \texttimes{} 6 \texttimes{} 1 mesh
of k-points. The calculation were performed within GGA-PBE approximation.

The alignment of Tc molecules on the silver (111) surface in $\gamma$-phase
and its surface unit cell are determined in accordance with the experimental
data \cite{soubatch_structure_2011}. The unit cell characterized
by the structural matrix $\left(\begin{array}{cc}
3 & 2\\
-5 & 2
\end{array}\right)$ contains only one molecule with the coverage of 83\%, being commensurate
with hexagonal lattice of Ag(111). The orientation of the molecular
longitudinal axis coincides with $[01\bar{1}]$-directions of the
silver substrate. To find the most favorable structure, four different
adsorption sites are used as a starting location of the Tc center
(see Fig. \ref{fig:tc-geo}) in the procedure of interatomic forces
minimization. The on-top $\Gamma_{0}$ position of Tc was proposed
in the experiment as the most probable one \cite{soubatch_structure_2011}.
Also, the bridge ($\Gamma_{2}$), hollow ($\Gamma_{3}$) and long
bridge ($\Gamma_{1}$) adsorption sites are considered (Fig. \ref{fig:tc-geo}). 

The unit cell of the $\alpha$-phase contains one Tc molecule with
the coverage of 100\%, which has the point-on-line coincidence with
the substrate and is described by the non-integer matrix $\left(\begin{array}{cc}
3.1 & 2\\
-2.25 & 3
\end{array}\right)$ \cite{langner_structural_2005,soubatch_structure_2011}. The commensurate
unit cell can be obtained by a tenfold increase of the first unit
vector and fourfold of the second one. As a result, the unit cell
contains already 40 molecules with their own adsorption sites. By
neglecting the shifts of the adsorption site along the first unit
vector (it is only 0.1 of the silver interatomic distance), the unit
cell can be tenfold shortened in this direction. Such a reduced unit
cell contains only four molecules on different adsorption sites and
has the 101\% coverage. This adapted unit cell is used in our calculations
of the electronic structure of the $\alpha$-Tc/Ag(111) interface,
which are performed within the Model 2 only (see Fig.\ref{fig:tc-geo}). 

The binding energy is introduced in the conventional way as the difference
between the total energy of the whole system $E_{sys}$ and the sum
of the energies of independently optimized freestanding monolayer
$E_{ml}$ and silver substrate $E_{sub}$. The counterpoise correction
\cite{boys_calculation_1970} is exploited to estimate the basis set
superposition error. To this end, the binding energy is written as
follows:

\begin{equation}
E_{b}^{(cp)}=E_{int}^{(cp)}-E_{def}^{ml}-E_{def}^{sub}\,.\label{eq:Eb-1}
\end{equation}

Here, the substrate-adsorbate interaction energy is expressed as $E_{int}^{(cp)}=E_{sys}-E_{ml}^{\ast(cp)}-E_{sub}^{\ast(cp)}$,
where both constituents of the system can be counterpoise corrected
(as indicated by the superscript in parentheses) or not, and they
both are cut from optimized geometry of the whole system. The deformation
term is simply the difference between the energy of the optimized
freestanding monolayer (or substrate) $E_{ml(sub)}$ and the energy
of the monolayer (substrate) $E_{ml(sub)}^{\ast}$ clamped in the
adsorption geometry.

\section{Results}

The calculated binding energy of $\gamma$-Tc/Ag(111) listed in Table
\ref{tab:Binding-energies} evidences that Tc molecules prefer to
be lined up directly above the silver rows by their external carbon
atoms, i. e., by those linked with hydrogens (Fig. \ref{fig:tc-geo}).
The long bridge ($\Gamma_{1}$) and on-top adsorption geometries ($\Gamma_{0}$)
provide disadvantageous alignments of Tc molecules on Ag(111); in
both cases carbons are located upon the interstitial sites of the
underlying silver layer (Fig. \ref{fig:tc-geo}). The same trends
in adsorption geometry were revealed in the recent calculation of
the Tc/Ag(110) interface \cite{tao_electronic_2015}. Because of the
obvious disadvantage of the $\Gamma_{1}$ geometry, it will excluded
from our further analysis, but despite the unfavorable adsorption
structure of $\Gamma_{0}$, it will be considered as a reference structure
proposed by experimentalists \cite{soubatch_structure_2011}. The
bridge ($\Gamma_{2}$) and hollow ($\Gamma_{3}$) sites are the most
favorable adsorption sites of $\gamma$-Tc and they both have very
close binding energies (Table \ref{tab:Binding-energies}). The above
statement holds true irrespectively of the exchange-correlation approximation
and type of the substrate model used.

\begin{table*}
\caption{\label{tab:Binding-energies} Vertical adsorption distances (in Å)
and binding energies (in eV) of $\gamma$-Tc/Ag(111) for different
models of the substrate. The distance is represented by the average
value of the normal distribution of vertical separations for carbon
and hydrogen atoms. The lattice constant of the silver substrate (in
Å) for every model is also given in square brackets. The superscript
in parenthesis denotes the spread of the vertical separations for
different C or H atoms in the ML, reflecting the distortion of the
Tc molecule deposited on top of the Ag(111) surface.}

\begin{ruledtabular}

\begin{tabular}{cccccccccccc}
 & $d_{z}$ (C)  & $d_{z}$ (H) & $E_{b}$ &  & $d_{z}$ (C) & $d_{z}$ (H) & $E_{b}$ &  & $d_{z}$ (C) & $d_{z}$ (H) & $E_{b}$\tabularnewline
\hline 
 & \multicolumn{3}{c}{top ($\Gamma_{0}$)} &  & \multicolumn{3}{c}{bridge ($\Gamma_{2}$)} &  & \multicolumn{3}{c}{hollow ($\Gamma_{3}$)}\tabularnewline
\textbf{GGA-PBE} &  &  &  &  &  &  &  &  &  &  & \tabularnewline
Model 1 {[}4.09{]} & $3.60^{(0.07)}$ & $3.57^{(0.09)}$ & -0.40 &  & $3.52^{(0.09)}$ & $3.48^{(0.10)}$ & -0.42 &  & $3.51^{(0.14)}$ & $3.45^{(0.23)}$ & -0.42\tabularnewline
Model 2 {[}4.16{]} & $3.57^{(0.09)}$ & $3.52^{(0.10)}$ & -0.48 &  & $3.49^{(0.15)}$ & $3.44^{(0.20)}$ & -0.50 &  & $3.41^{(0.15)}$ & $3.37^{(0.21)}$ & -0.49\tabularnewline
Model 3 {[}4.23{]} & $3.41^{(0.02)}$ & $3.38^{(0.03)}$ & -0.32 &  & $3.32^{(0.08)}$ & $3.27^{(0.13)}$ & -0.35 &  & $3.32^{(0.11)}$ & $3.27^{(0.18)}$ & -0.35\tabularnewline
\textbf{optB88-vdW} &  &  &  &  &  &  &  &  &  &  & \tabularnewline
Model 1 {[}4.09{]} & $3.22^{(0.02)}$ & $3.19^{(0.01)}$ & -1.96 &  & $3.14^{(0.04)}$ & $3.11^{(0.06)}$ & -2.05 &  & $3.16^{(0.08)}$ & $3.13^{(0.17)}$ & -2.05\tabularnewline
Model 2 {[}4.17{]} & $3.21^{(0.03)}$ & $3.19^{(0.03)}$ & -1.97 &  & $3.16^{(0.07)}$ & $3.13^{(0.12)}$ & -2.06 &  & $3.15^{(0.10)}$ & $3.14^{(0.18)}$ & -2.06\tabularnewline
Model 3 {[}4.23{]} & $3.10^{(0.04)}$ & $3.09^{(0.04)}$ & -1.86 &  & $3.00^{(0.05)}$ & $2.98^{(0.08)}$ & -1.97 &  & $3.02^{(0.06)}$ & $3.01^{(0.11)}$ & -1.97\tabularnewline
\end{tabular}\end{ruledtabular}
\end{table*}

\subsection{Adsorption geometry}

For the same adsorption site, the vertical separation of carbon atoms
as found with the GGA-PBE functional within Model 1 and 2 is characterized
by the fairly close averaged values (see Table \ref{tab:Binding-energies});
the difference is merely of 0.03~Å in the case of the on-top and
bridge positions of the molecule. Model 3 gives the noticeably smaller
adsorption height for all adsorption sites. It is worth also noting
that the Tc molecule is rather distorted on the Ag(111) surface. Actually,
the \emph{z}-coordinate of the molecule atoms varies within the range
of 0.07--0.1~Å. It is clearly seen in Table \ref{tab:Binding-energies},
where the superscript in parentheses denotes the maximal difference
between the perfect \emph{z}-coordinate of the topmost silver layer
and the \emph{z}-position of carbon and hydrogen atoms of the molecule.
For the on-top adsorption geometry, the distortion is found along
the longitudinal axis only, and the center of the molecule is farther
from the substrate than the molecule edges. The attractive interaction
between carbon and underlying silver atoms results in more complicated
distortion of Tc, when the center of the latter occupies the bridge
or hollow site. For instance, the lateral slop of Tc in the hollow
($\Gamma_{3}$) site takes place, i.e., the side of the molecule,
on which the external carbon atoms locate directly over the silver
atoms is about 0.08~Å closer to the substrate than the side with
carbon atoms located over the interstitials (see Fig. \ref{fig:tc-geo}). 

The inclusion of the long range interaction within optB88-vdW approximation
gives rise to a reduction of the vertical adsorption distance (Table
\ref{tab:Binding-energies}). Model 1 and 2 provide almost identical
vertical separations and binding energies. As it was before, Model
3 using the enlarged Ag lattice constant results in shorter separations
and a weaker binding. The adsorption geometry $\Gamma_{0}$ is characterized
by the nearly flat shape of Tc, while, as in the previous case, the
molecule is more distorted in $\Gamma_{2}$ and $\Gamma_{3}$ in accordance
with the mutual alignment of carbon and silver atoms. Moreover, in
the case of the hollow adsorption site the slop of the Tc molecule
has the same value regardless the approximation chosen for the exchange-correlation
functional.

The experimental estimation of the binding energy for $\alpha$-Tc
by the thermal desorption study \cite{gonella_tetracene_2008} is
about 1~eV (1.4~eV in the zero-coverage limit), while for $\gamma$-Tc
it is expected to be a slightly higher, because of the smaller coverage.
The calculated binding energy of $\Gamma_{2}$ structure is thus considerably
underestimated within the GGA-PBE calculations (Table \ref{tab:Binding-energies});
as compared with the experimental value the energy of 0.5~eV is half
as much. On the contrary, the calculation with the optB88-vdW functional
provides $E_{b}$, which is twice as much as compared with the experiment
(Table \ref{tab:Binding-energies}). 

The detailed insight into the substrate-adsorbate interaction is done
for the bridge ($\Gamma_{2}$) geometry, because it is one of the
most energetically favorable structure. The major part of the binding
energy is the interaction energy (Table \ref{tab:Gamma2}), while
the deformation energy of both the molecular monolayer and the topmost
layers of the substrate is of $\sim$20 meV in the PBE calculations
and slightly larger in the case of the optB88-vdW functional. As compared
with the GGA-PBE values, the long-range dispersive interaction gives
rise to a fourfold increase in the interaction energy. 

\begin{table*}
\protect\caption{\label{tab:Gamma2} Energy terms (eV) of equation \ref{eq:Eb-1} (with
and without the BSSE correction), the energies of the former frontier
orbitals of the Tc molecules (eV) as measured from the Fermi energy
of the interface, and the amount of charge (\emph{-e}) in the molecular
monolayer region for the bridge adsorption geometry $\Gamma_{2}$
of the $\gamma$ and $\alpha$-Tc/Ag(111) interfaces. }

\begin{ruledtabular}

\begin{tabular*}{0.9\textwidth}{@{\extracolsep{\fill}}rcccccc}
 & $d_{z}$ & $E_{int}/E_{int}^{cp}$ & $E_{def}^{ml}$ & $E_{def}^{sub}$ & $E_{b}/E_{b}^{cp}$ & $Q(z_{0})/Q^{cp}(z_{0})$\tabularnewline
\hline 
 & \textbf{$\gamma$-Tc} &  &  &  &  & \tabularnewline
\textbf{GGA-PBE} &  &  &  &  &  & \tabularnewline
Model 1 {[}4.09{]} & 3.52 & -0.43/-0.02 & 0.00 & \-\-\- & -0.42/-0.01 & 0.10/0.14\tabularnewline
Model 2 {[}4.16{]} & 3.49 & -0.53/-0.02 & -0.01 & -0.01 & -0.50/0.00 & 0.07/0.12\tabularnewline
Model 3 {[}4.23{]} & 3.32 & -0.37/-0.01 & -0.01 & -0.02 & -0.35/0.01 & 0.12/0.16\tabularnewline
\textbf{optB88-vdW} &  &  &  &  &  & \tabularnewline
Model 1 {[}4.09{]} & 3.14 & -2.06/-1.50 & -0.01 & \-\-\- & -2.05/-1.49 & 0.13/0.20\tabularnewline
Model 2 {[}4.17{]} & 3.16 & -2.09/-1.51 & -0.01 & -0.01 & -2.06/-1.48 & 0.09/0.18\tabularnewline
Model 3 {[}4.23{]} & 3.00 & -2.01/-1.54 & -0.01 & -0.03 & -1.97/-1.50 & 0.14/0.21\tabularnewline
 & \textbf{$\alpha$-Tc} &  &  &  &  & \tabularnewline
\textbf{optB88-vdW} &  &  &  &  &  & \tabularnewline
Model 2 {[}4.17{]} & 3.16 & -1.87/-1.44 & -0.01 & -0.02 & -1.84/-1.41 & 0.10/0.16\tabularnewline
\end{tabular*}

\end{ruledtabular}
\end{table*}

The binding energy computed with the use of localized basis functions
tends to be overestimated by reason of BSSE, which should be corrected.
There is a noticeable reduction in counterpoise corrected substrate-adsorbate
interaction, because of quite large area of Tc molecule for both functionals
used. In the GGA-PBE calculations, the considered $\Gamma_{2}$ structure
becomes even unstable (Table \ref{tab:Gamma2}), being characterized
by the positive binding energy. In the case of the optB88-vdW functional,
the counterpoise corrected binding energy comes closer to the experimental
value \cite{gonella_tetracene_2008}, but it is still overestimated.
Thus, the binding of the molecular monolayer with silver in $\gamma$-Tc/Ag(111)
is fully determined by long-range dispersive forces. Note also that
the BSSE-free values of the binding energy are almost the same for
all models of the substrate; its deviation is about 20~meV.

The calculation of $\alpha$-Tc/Ag(111) was performed in Model 2 with
optB88-vdW functional only. As was described above, the coverage in
adapted unit cell coincides well with the experimental one. The adsorption
sites for Tc were chosen in accordance with $\gamma$-phase calculation,
i.e., the external carbon atoms reside above the silver atoms (Fig.
\ref{fig:tc-geo}) as in the bridge and hollow adsorption sites. The
averaged vertical separations for carbon (3.16~Å) and hydrogen (3.13~Å)
atoms are very close to the $\gamma$-phase results (see table \ref{tab:Gamma2}),
but the binding energy becomes smaller as compared with the $\gamma$-phase
because of the increased coverage.

\subsection{Charge transfer and bands alignment}

Due to the substrate-adsorbate interaction, the real space electron
density rearrangement occurs in the interface region. The charge transfer
between the substrate and the molecular monolayer can be extracted
from the laterally averaged electron-density difference, which is
calculated in the following way:

\[
\Delta n(z)=n_{sys}(z)-n_{ml}(z)-n_{sub}(z)\,.
\]

\begin{figure}
\includegraphics[width=0.99\columnwidth]{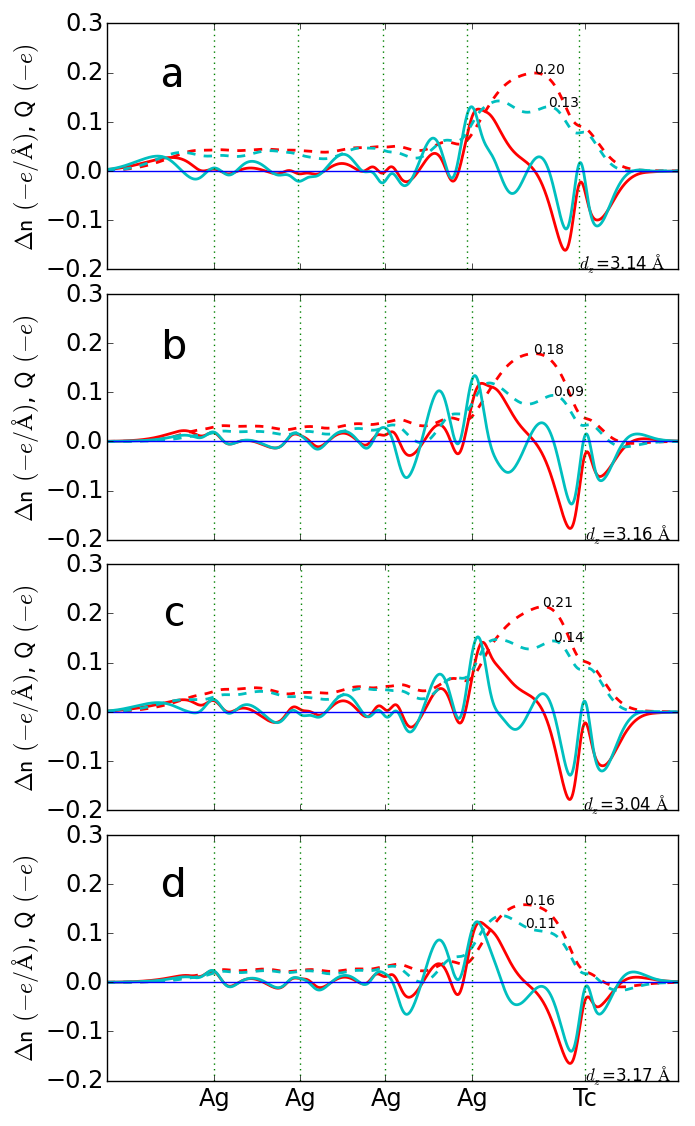}

\protect\caption{\label{fig:Charge-transfer} The charge density difference $\Delta n(z)$
averaged in \emph{xy}-plane (solid lines) and the respective amount
of charge transfer $Q(z)$ (dashed lines) with (red) and without (cyan)
the BSSE correction for the bridge ($\Gamma_{2}$) adsorption site.
a) Model 1, b) Model 2, c) Model 3 of $\gamma$-Tc and d) Model 2
of $\alpha$-Tc}
\end{figure}

Here, $n_{sys}$ is the charge density of the interface under study
averaged over the $xy$-plane within the unit cell, $n_{ml}$ and
$n_{sub}$ are the charge densities of the monolayer and the substrate,
respectively. A positive value of $\Delta n(z)$ indicates a gain
in electron density upon adsorption, while a negative value means
a loss of electrons. In Fig. \ref{fig:Charge-transfer}, it is clearly
seen that the regions around molecular plane loose electrons. The
amount of change transfer can be thus determined by function $Q(z)=\int_{z_{0}}^{z}\Delta n(z')\, dz'$,
here $z_{vac}$ is a point in the vacuum
where $\Delta n(z)$ is zero. Since the integration is started from
the clean side of the slab, which is on the left in Fig. \ref{fig:Charge-transfer},
the positive values of $Q(z)$ at a given $z$ provide the amount
of charge, which flow from the right side into the left side with
respect to this $z$ and vise versa for the negative values. If we
fix the boundary between substrate and adsorbate at the point in interface
region, where $\Delta n(z)$ crosses zero on the left of its global
minimum, then the positive $Q(z_{0})$ will determine the charge transfer
from molecule to metal. The charges presented in Table \ref{tab:Gamma2}
have values which are close to each other. Nevertheless, $Q(z_{0})$
distinctly depends on adsorption distance; it tends to be increased
with shortening of the molecule-substrate separation. Additionally,
$\Delta n$ is sensitive to the overlap between the molecular and
the substrate wave functions, it is clearly seen on example of Model
2 (see Fig. \ref{fig:Charge-transfer} b, d), where the basis functions
of the topmost silver atoms with increased $r_{c}$ were used. The
influence of the wave functions overlap on the amount of charge flow
is also observed in the case of the counterpoise corrected charge
difference. The superposition of substrate and molecule basis functions
is expected to cause an error in a charge density in the same way
as in the interaction energy. The amount of charge transfered into
the substrate noticeably increased due to the redistribution of BSSE
free charge density (see Table \ref{tab:Gamma2}), and the accumulation
of charge density upon silver layer becomes evident (see Fig. \ref{fig:Charge-transfer}).

The aforementioned behavior of $\Delta n(z)$ and $Q(z)$ can be interpreted
in terms of the push-back effect \cite{witte_vacuum_2005}, which
is typical for organic molecules physisorbed onto metallic surfaces.
Actually, the repulsive exchange interaction between the molecular
and substrate electrons, i.e., Pauli repulsion, leads to the molecular
electronic cloud pushes the substrate electronic cloud back into the
metal. The same mechanism is responsible for charge transfer in the
repeatedly described PTCDA/Au(111) interface \cite{romaner_theoretical_2009,rusu_first-principles_2010}. 

The charge density difference of the $\alpha$-phase with increased
coverage of Tc on Ag(111) provides almost the same outflow of charge
from the ML as in the $\gamma$-phase, i.e. the increase of the charge
density in the ML has a minimal effect on the amount of charge transfered.
Note, a negative work function change upon adsorption of Tc has been
experimentally observed \cite{frank_unoccupied_1988}, suggesting
donation of negative charge from the molecule to the Ag substrate,
which was roughly estimated afterwards \cite{gonella_tetracene_2008};
the present calculation is thus consistent with experimental findings.

\begin{figure*}
\includegraphics[width=0.98\textwidth]{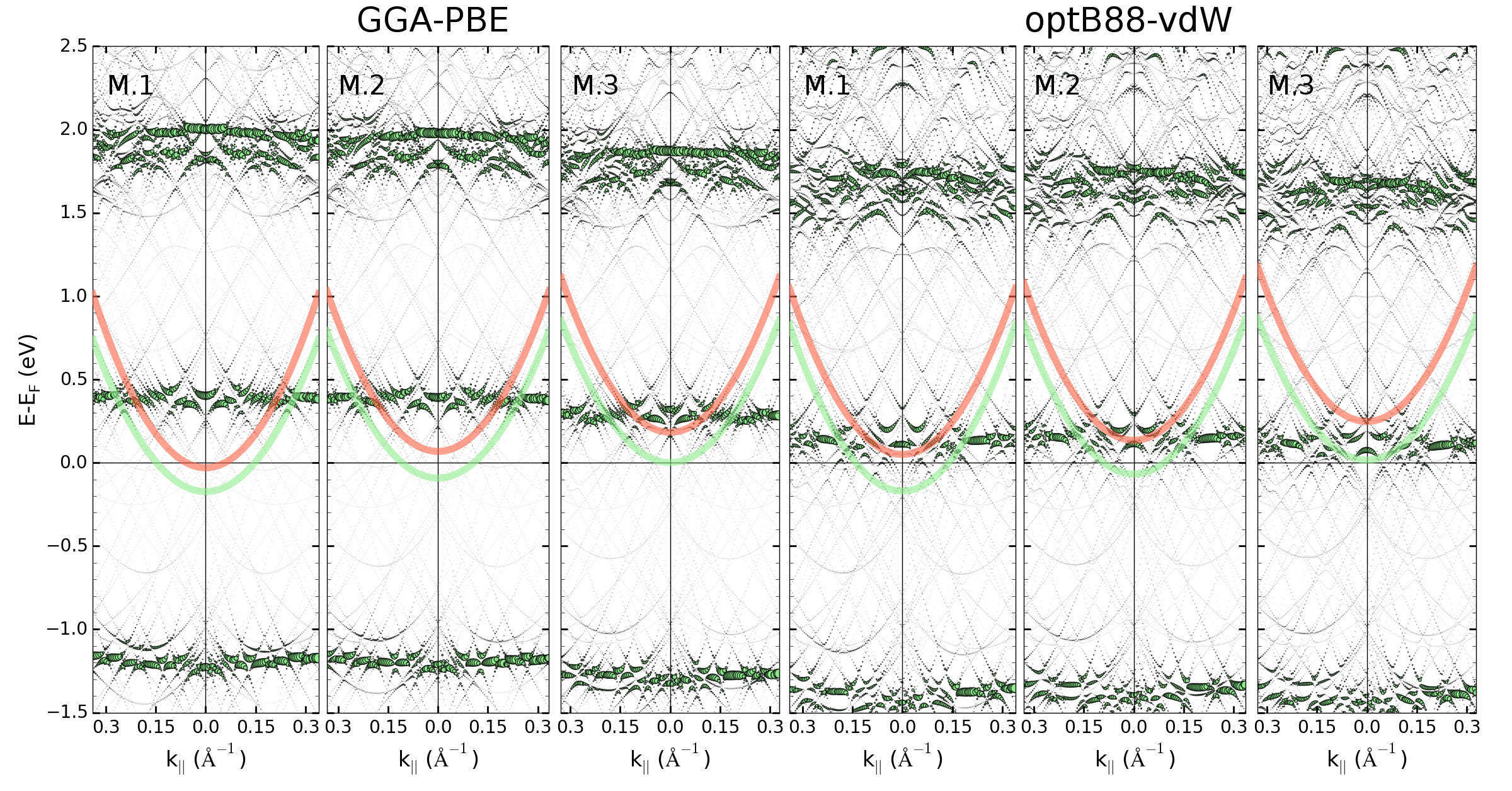}

\protect\caption{\label{fig:Projected-band-structure}Band structure of the $\gamma$-Tc/Ag(111)
interface projected onto the \emph{$p{}_{z}$} molecular states. The
dispersive interface (red thick line) and surface (green) states are
presented as well.}
\end{figure*}

The fact that the molecular ML looses electrons implies that there
is an electron donation from Tc to the substrate involving many formerly
occupied molecular orbitals and that the LUMO (the lowest unoccupied
molecular orbital) of Tc is still empty after the interaction with
silver. The projected band structure onto $p_{z}$-states of carbon
atoms in the directions of the reciprocal lattice vectors was calculated
for the system where silver substrate is represented by a 12-layer
slab ( Fig. \ref{fig:Projected-band-structure}). As clearly seen
in the figure, the energy-bands alignment of the interacting systems
with respect to the Fermi level ensures that the LUMO of Tc remains
unoccupied and non-dispersive, residing close to the Fermi level,
especially in the optB88-vdW calculations. 

The band structure calculated within Model 1 and 2 shows similar energy
values for the unoccupied LUMO and LUMO+1 (Table \ref{tab:IS}), while
these orbitals are shifted down in the case of Model 3 for both functionals
(see Table \ref{tab:Gamma2}). This down-shift can be explained by
the decrease of the substrate-adsorbate distance. Therefore, the calculated
energy of the LUMO with the GGA-PBE functional (390 meV) is higher
than the value of 150 meV obtained with the optB88-vdW one (Table
\ref{tab:IS}). Since the adsorption height of the$\alpha$-phase
considered in Model 2 within optB88-vdW approximation is the same
as of the $\gamma$-phase, the energy of the LUMO has close value
too, but now this band is characterized by a larger width.

The adsorption of the Tc monolayer on Ag(111) in the $\gamma$-phase
results in an up-shift of the silver surface state (SS). The same
transformation of SS manifests itself after adsorption of the NTCDA
and PTCDA \cite{zaitsev_change_2010,marks_energy_2011,zaitsev_transformation_2012}
or phthalocyanine \cite{caplins_metal/phthalocyanine_2014} molecules
on the same substrate, whereas the downshift of the SS was observed
for pentacene molecules absorbed on Cu(110) \cite{scheybal_modification_2009}.
The projected band structure onto the $p_{z}$-states of silver atoms
located in the uppermost substrate layer covered by the ML or the
clean lowermost silver layer directly reveals the interface state
(IS) or the clean-side SS of the slab, respectively. Both states are
shown in Fig. \ref{fig:Projected-band-structure} by red (IS) and
green (SS) dispersive curves. 

The energies of both states at the $\bar{\Gamma}$-point depend strongly
on the type of the substrate model. The calculated SS energy of the
clean Ag(111) surface in the slab model is very sensitive to the silver
lattice constant; the greater the equilibrium lattice constant that
the bulk silver possesses the higher the SS energy we obtain. The
same dependence is observed in the behavior of both the clean-side
SS of the $\gamma$-Tc/Ag(111) slab and the IS. It is worth noting
that if an extremely thick slab is used, the bonding-antibonding splitting
of two surface states (since we have two surfaces of the slab) would
be negligible. After the adsorption of the molecular monolayer on
one side of the slab, the respective SS transforms into the interface
state, while the other tends to be a surface state of the thick slab
limit. Therefore, the difference between the resulting IS and SS represents
the surface-state transformation energy. As follows from Table \ref{tab:IS},
in Model 1 and 2 the differences for the $\gamma$-Tc/Ag(111) interface
are fairly close, while in Model 3 it has a larger value due to the
smaller adsorption distance \cite{zaitsev_change_2010,zaitsev_transformation_2012}.
Definitely, the GGA-PBE calculation provides smaller IS energies then
optB88-vdW one for the same reason.

\begin{figure}
\includegraphics[width=1\columnwidth]{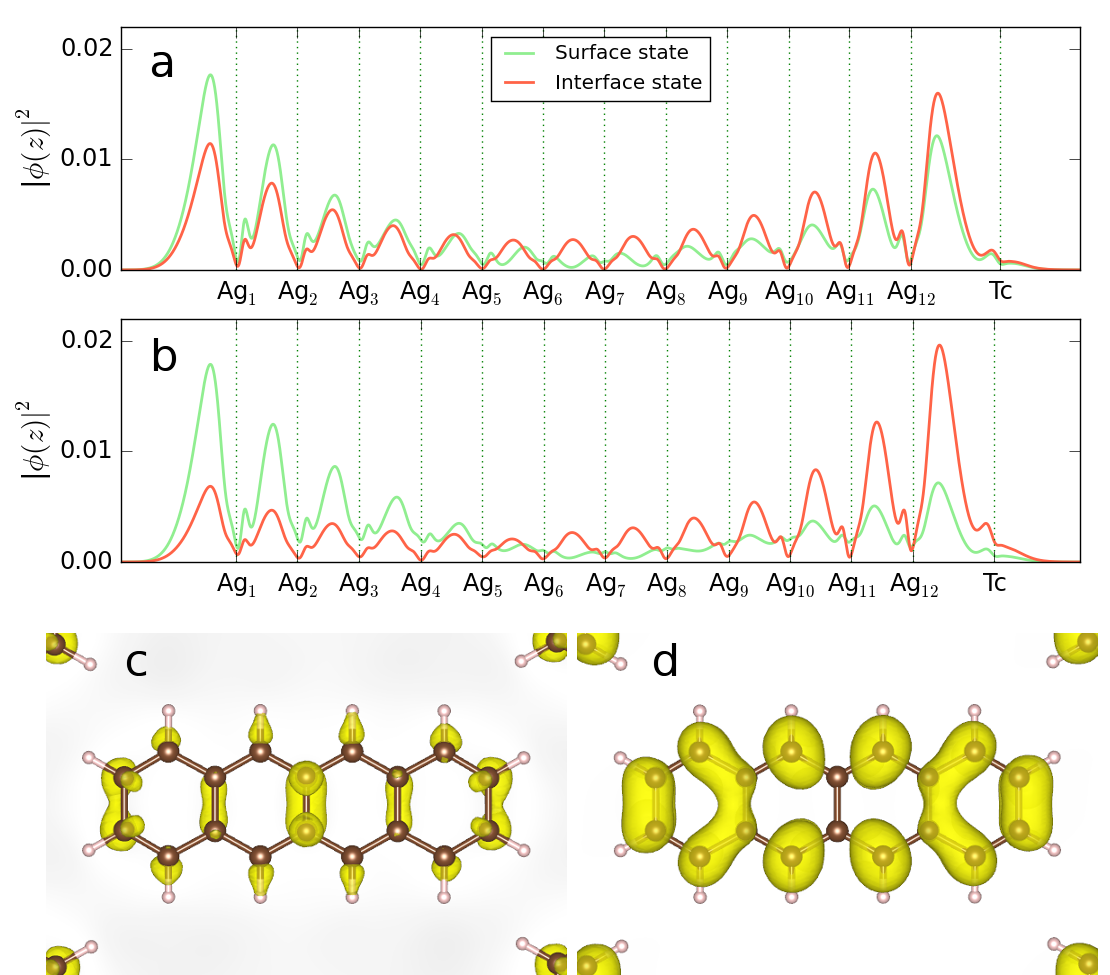}

\protect\caption{\label{fig:rho_z_IS} Charge density distribution averaged over the
spatial $xy$-coordinates within the unit cell of the $\gamma$-Tc/Ag(111)
for adsorption geometries obtained within Model 2 with (a) GGA-PBE
and (b) optB88-vdW . The interface (red) and clean-side surface state
(green) of the slab at the $\bar{\Gamma}$-point are presented as
functions of $z$, and their dispersion is shown on figure \ref{fig:Projected-band-structure}.
(c) Charge-density isosurfaces of the IS and (d) the former LUMO.}
\end{figure}

The adsorption distance is responsible for the spatial localization
of the IS wave function. The perturbation of the SS by adsorbed molecules
becomes stronger with decreasing the distance. For the quite big adsorption
distance obtained in the GGA-PBE calculation, the charge density of
the IS is localized almost equally on both sides of the slab as well
as the clean-side SS (see Fig. \ref{fig:rho_z_IS} a). For the shorter
distance obtained with optB88-vdW functional, the IS tends to be more
localized in the interface region, while the SS on the clean side
(Fig \ref{fig:rho_z_IS} b). In the case of NTCDA and PTCDA on Ag(111),
there is a strong chemical interaction between the functional group
and metallic atoms, therefore the adsorption distance is much shorter,
the perturbation of the former SS is stronger, so the IS wave function
is almost completely localized on the side with the adsorbate, whereas
the SS resides on the opposite clean side of the slab \cite{zaitsev_transformation_2012}.

The overlap between the IS and molecular state wave functions exhibits
the same dependence from the substrate-adsorbate vertical separation.
For the largest distance, the accumulation of the IS charge density
on the molecular ML has a minimal value, but it increases as the adsorbate
approaches the substrate. The shape of the IS charge density localized
on the Tc molecule (Fig. \ref{fig:rho_z_IS} c) is insensitive to
the adsorption distance, and there is no hybridization of the IS with
the unoccupied former LUMO of Tc (Fig. \ref{fig:rho_z_IS}d) though
they have close energies at the $\bar{\Gamma}$-point. In contrast,
in the strongly interacting systems as NTCDA or PTCDA on Ag(111),
the former LUMO is partially occupied and the shape of the IS in molecular
region resembles the LUMO of the free molecule.

Another factor affecting the magnitude of the SS transformation is
the adsorbate coverage or the density of carbon atoms in the surface
unit cell \cite{zaitsev_transformation_2012}. The $\alpha$-phase
calculations illustrate the influence of the increased coverage by
$\sim18\%$; the transformation of the SS-IS energy is increased by
30 meV in comparison with the $\gamma$-phase calculations within
Model 2 (Table \ref{tab:IS}). The available STS (scanning tunneling
spectroscopic) data \cite{soubatch_structure_2011} of $\alpha$-Tc/Ag(111)
provides the SS-IS energy difference of 220 meV, which is well described
by the optB88-vdW calculation. 

\begin{table}
\protect\caption{\label{tab:IS}The energies (in meV) of the interface state ($E_{IS}$)
and the surface state ($E_{SS}$) and their difference $\delta E$
for the bridge adsorption position. Tc/Ag(111) silver slab has the
12-layer thickness.}

\begin{ruledtabular}

\begin{tabular}{cccccccccc}
 & $E_{IS}$ & $E_{SS}$ & $\delta E$ & $E_{LUMO}$ &  & $E_{IS}$ & $E_{SS}$ & $\delta E$ & $E_{LUMO}$\tabularnewline
\hline 
 & \multicolumn{3}{c}{\textbf{GGA-PBE}} &  &  & \multicolumn{3}{c}{\textbf{optB88-vdW}} & \tabularnewline
\textbf{$\gamma$-Tc} &  &  &  &  &  &  &  &  & \tabularnewline
Model 1 & -20 & -172 & 152 & 390 &  & 50 & -164 & 214 & 140\tabularnewline
Model 2 & 68 & -92 & 160 & 390 &  & 135 & -70 & 205 & 150\tabularnewline
Model 3 & 183 & 1 & 182 & 280 &  & 248 & 16 & 232 & 110\tabularnewline
\textbf{$\alpha$-Tc} &  &  &  &  &  &  &  &  & \tabularnewline
Model 2 &  &  &  &  &  & 175 & -60 & 235 & 150\tabularnewline
exp \cite{soubatch_structure_2011} &  &  &  &  &  &  &  & 220 & 830\tabularnewline
\end{tabular}

\end{ruledtabular}
\end{table}

\section{Conclusion}

We performed the theoretical study of the Tc/Ag(111) metal-organic
interface by means of DFT calculations with localized basis. We focused
on the simultaneous description of the structural and electronic properties
on equal footings. The optB88-vdW version of the vdW-DF2 family of
exchange-correlation functionals especially optimized for solids was
used as well as the conventional GGA-PBE approximation. 

The bridge and hollow adsorption sites were determined as the most
stable geometries of $\gamma$-Tc/Ag(111) irrespective to the approximation
chosen for the exchange-correlation functional, but vdW contribution
constitutes a decisive part of the substrate-adsorbate interaction.
The inclusion of the long-range dispersive forces into the calculations
provides the close agreement of the binding energy with the available
experimental data \cite{gonella_tetracene_2008} and, therefore, better
description of the adsorption geometry, whereas the conventional GGA-PBE
scheme underestimates the equilibrium adsorption distance.

We suggested and tested three schemes of handling the substrate in
the metal-organic hybrid interface under study in order to overcome
the controversial trends of its bulk and surface electronic properties
in the case of localized-basis-set description. We obtained the best
result within the Model 2, where only the silver atoms on the slab
surfaces are described by long-range basis orbitals, while the internal
(bulk) silver atoms are characterized by rather short orbitals providing
adequate bulk properties. In our case of weakly bound adsorbate, the
corrugation of the substrate is small and the Model 1 with fixed substrate
provides a good description as well. But when considering chemisorbed
systems, like PTCDA/Ag(111), the stronger substrate-adsorbate interaction
gives rise to the larger substrate corrugation and the possibility
for the substrate to be relaxed becomes quite important. 

We showed that too large lattice constant of the substrate (Model
3) results in noticeable change of adsorption geometry and, in consequence,
to inadequate description of the electronic structure. Moreover, the
absolute position of the interface state (or surface state in case
of pristine surface) is systematically shifted upward with growth
of the substrate lattice constant. But, the difference between the
interface and surface state energies is tolerant to the substrate
characteristics, it is sensitive only to the adsorption distance of
the adsorbate. The interface states of the both phases are fully unoccupied,
and because of the enlarged density of Tc molecules in the $\alpha$-phase
the IS energy is higher than in the $\gamma$-phase. The calculation
with the OptB88-vdW functional gives not only good value of the binding
energy, but the IS energy is in close agreement with the experimental
measurements \cite{soubatch_structure_2011}.
\begin{acknowledgments}
This work is a project of the SFB 1083 ``Structure and Dynamics of
Internal Interfaces'' funded by the Deutsche Forschungsgemeinschaft
(DFG). We acknowledge partial support from the University of Basque
Country UPV/EHU (IT-756-13), the Departamento de Educaci\'{o}n del
Gobierno Vasco, The Tomsk State University Academic D.I. Mendeleev
Fund Program (Grant No. 8.1.05.2015), the Spanish Ministry of Economy
and Competitiveness MINECO (Grant No. FIS2013-48286-C2-1-P), and Saint
Petersburg State University (Project No. 11.50.202.2015).
\end{acknowledgments}

\bibliographystyle{apsrev4-1}
\bibliography{ref}

\begin{thebibliography}{49}%
\makeatletter
\providecommand \@ifxundefined [1]{%
 \@ifx{#1\undefined}
}%
\providecommand \@ifnum [1]{%
 \ifnum #1\expandafter \@firstoftwo
 \else \expandafter \@secondoftwo
 \fi
}%
\providecommand \@ifx [1]{%
 \ifx #1\expandafter \@firstoftwo
 \else \expandafter \@secondoftwo
 \fi
}%
\providecommand \natexlab [1]{#1}%
\providecommand \enquote  [1]{``#1''}%
\providecommand \bibnamefont  [1]{#1}%
\providecommand \bibfnamefont [1]{#1}%
\providecommand \citenamefont [1]{#1}%
\providecommand \href@noop [0]{\@secondoftwo}%
\providecommand \href [0]{\begingroup \@sanitize@url \@href}%
\providecommand \@href[1]{\@@startlink{#1}\@@href}%
\providecommand \@@href[1]{\endgroup#1\@@endlink}%
\providecommand \@sanitize@url [0]{\catcode `\\12\catcode `\$12\catcode
  `\&12\catcode `\#12\catcode `\^12\catcode `\_12\catcode `\%12\relax}%
\providecommand \@@startlink[1]{}%
\providecommand \@@endlink[0]{}%
\providecommand \url  [0]{\begingroup\@sanitize@url \@url }%
\providecommand \@url [1]{\endgroup\@href {#1}{\urlprefix }}%
\providecommand \urlprefix  [0]{URL }%
\providecommand \Eprint [0]{\href }%
\providecommand \doibase [0]{http://dx.doi.org/}%
\providecommand \selectlanguage [0]{\@gobble}%
\providecommand \bibinfo  [0]{\@secondoftwo}%
\providecommand \bibfield  [0]{\@secondoftwo}%
\providecommand \translation [1]{[#1]}%
\providecommand \BibitemOpen [0]{}%
\providecommand \bibitemStop [0]{}%
\providecommand \bibitemNoStop [0]{.\EOS\space}%
\providecommand \EOS [0]{\spacefactor3000\relax}%
\providecommand \BibitemShut  [1]{\csname bibitem#1\endcsname}%
\let\auto@bib@innerbib\@empty
\bibitem [{\citenamefont {Dimitrakopoulos}\ and\ \citenamefont
  {Malenfant}(2002)}]{dimitrakopoulos_organic_2002}%
  \BibitemOpen
  \bibfield  {author} {\bibinfo {author} {\bibfnamefont {C.}~\bibnamefont
  {Dimitrakopoulos}}\ and\ \bibinfo {author} {\bibfnamefont {P.}~\bibnamefont
  {Malenfant}},\ }\href@noop {} {\bibfield  {journal} {\bibinfo  {journal}
  {Advanced Materials}\ }\textbf {\bibinfo {volume} {14}},\ \bibinfo {pages}
  {99} (\bibinfo {year} {2002})}\BibitemShut {NoStop}%
\bibitem [{\citenamefont {Hoppe}\ and\ \citenamefont
  {Sariciftci}(2004)}]{hoppe_organic_2004}%
  \BibitemOpen
  \bibfield  {author} {\bibinfo {author} {\bibfnamefont {H.}~\bibnamefont
  {Hoppe}}\ and\ \bibinfo {author} {\bibfnamefont {N.~S.}\ \bibnamefont
  {Sariciftci}},\ }\href@noop {} {\bibfield  {journal} {\bibinfo  {journal}
  {Journal of Materials Research}\ }\textbf {\bibinfo {volume} {19}},\ \bibinfo
  {pages} {1924} (\bibinfo {year} {2004})}\BibitemShut {NoStop}%
\bibitem [{\citenamefont {Shen}\ \emph {et~al.}(2004)\citenamefont {Shen},
  \citenamefont {Hosseini}, \citenamefont {Wong},\ and\ \citenamefont
  {Malliaras}}]{shen_how_2004}%
  \BibitemOpen
  \bibfield  {author} {\bibinfo {author} {\bibfnamefont {Y.}~\bibnamefont
  {Shen}}, \bibinfo {author} {\bibfnamefont {A.~R.}\ \bibnamefont {Hosseini}},
  \bibinfo {author} {\bibfnamefont {M.~H.}\ \bibnamefont {Wong}}, \ and\
  \bibinfo {author} {\bibfnamefont {G.~G.}\ \bibnamefont {Malliaras}},\
  }\href@noop {} {\bibfield  {journal} {\bibinfo  {journal} {ChemPhysChem}\
  }\textbf {\bibinfo {volume} {5}},\ \bibinfo {pages} {16} (\bibinfo {year}
  {2004})}\BibitemShut {NoStop}%
\bibitem [{\citenamefont {Koch}\ \emph {et~al.}(2013)\citenamefont {Koch},
  \citenamefont {Ueno},\ and\ \citenamefont
  {Wee~(Eds.)}}]{koch_moleculemetal_2013}%
  \BibitemOpen
  \bibfield  {author} {\bibinfo {author} {\bibfnamefont {N.}~\bibnamefont
  {Koch}}, \bibinfo {author} {\bibfnamefont {U.}~\bibnamefont {Ueno}}, \ and\
  \bibinfo {author} {\bibfnamefont {A.}~\bibnamefont {Wee~(Eds.)}},\
  }\href@noop {} {\emph {\bibinfo {title} {The Molecule-metal interface}}}\
  (\bibinfo  {publisher} {Wiley-VCH},\ \bibinfo {address} {Weinheim},\ \bibinfo
  {year} {2013})\BibitemShut {NoStop}%
\bibitem [{\citenamefont {Koch}(2008)}]{koch_energy_2008}%
  \BibitemOpen
  \bibfield  {author} {\bibinfo {author} {\bibfnamefont {N.}~\bibnamefont
  {Koch}},\ }\href@noop {} {\bibfield  {journal} {\bibinfo  {journal} {Journal
  of Physics: Condensed Matter}\ }\textbf {\bibinfo {volume} {20}},\ \bibinfo
  {pages} {184008} (\bibinfo {year} {2008})}\BibitemShut {NoStop}%
\bibitem [{\citenamefont {Tautz}(2007)}]{tautz_structure_2007}%
  \BibitemOpen
  \bibfield  {author} {\bibinfo {author} {\bibfnamefont {F.}~\bibnamefont
  {Tautz}},\ }\href@noop {} {\bibfield  {journal} {\bibinfo  {journal}
  {Progress in Surface Science}\ }\textbf {\bibinfo {volume} {82}},\ \bibinfo
  {pages} {479} (\bibinfo {year} {2007})}\BibitemShut {NoStop}%
\bibitem [{\citenamefont {Temirov}\ \emph {et~al.}(2006)\citenamefont
  {Temirov}, \citenamefont {Soubatch}, \citenamefont {Luican},\ and\
  \citenamefont {Tautz}}]{temirov_free-electron-like_2006}%
  \BibitemOpen
  \bibfield  {author} {\bibinfo {author} {\bibfnamefont {R.}~\bibnamefont
  {Temirov}}, \bibinfo {author} {\bibfnamefont {S.}~\bibnamefont {Soubatch}},
  \bibinfo {author} {\bibfnamefont {A.}~\bibnamefont {Luican}}, \ and\ \bibinfo
  {author} {\bibfnamefont {F.~S.}\ \bibnamefont {Tautz}},\ }\href@noop {}
  {\bibfield  {journal} {\bibinfo  {journal} {Nature}\ }\textbf {\bibinfo
  {volume} {444}},\ \bibinfo {pages} {350} (\bibinfo {year}
  {2006})}\BibitemShut {NoStop}%
\bibitem [{\citenamefont {Schwalb}\ \emph {et~al.}(2008)\citenamefont
  {Schwalb}, \citenamefont {Sachs}, \citenamefont {Marks}, \citenamefont
  {Sch\"{o}ll}, \citenamefont {Reinert}, \citenamefont {Umbach},\ and\
  \citenamefont {H\"{o}fer}}]{schwalb_electron_2008}%
  \BibitemOpen
  \bibfield  {author} {\bibinfo {author} {\bibfnamefont {C.~H.}\ \bibnamefont
  {Schwalb}}, \bibinfo {author} {\bibfnamefont {S.}~\bibnamefont {Sachs}},
  \bibinfo {author} {\bibfnamefont {M.}~\bibnamefont {Marks}}, \bibinfo
  {author} {\bibfnamefont {A.}~\bibnamefont {Sch\"{o}ll}}, \bibinfo {author}
  {\bibfnamefont {F.}~\bibnamefont {Reinert}}, \bibinfo {author} {\bibfnamefont
  {E.}~\bibnamefont {Umbach}}, \ and\ \bibinfo {author} {\bibfnamefont
  {U.}~\bibnamefont {H\"{o}fer}},\ }\href@noop {} {\bibfield  {journal}
  {\bibinfo  {journal} {Physical Review Letters}\ }\textbf {\bibinfo {volume}
  {101}},\ \bibinfo {pages} {146801} (\bibinfo {year} {2008})}\BibitemShut
  {NoStop}%
\bibitem [{\citenamefont {Zaitsev}\ \emph {et~al.}(2010)\citenamefont
  {Zaitsev}, \citenamefont {Nechaev},\ and\ \citenamefont
  {Chulkov}}]{zaitsev_change_2010}%
  \BibitemOpen
  \bibfield  {author} {\bibinfo {author} {\bibfnamefont {N.~L.}\ \bibnamefont
  {Zaitsev}}, \bibinfo {author} {\bibfnamefont {I.~A.}\ \bibnamefont
  {Nechaev}}, \ and\ \bibinfo {author} {\bibfnamefont {E.~V.}\ \bibnamefont
  {Chulkov}},\ }\href@noop {} {\bibfield  {journal} {\bibinfo  {journal}
  {Journal of Experimental and Theoretical Physics}\ }\textbf {\bibinfo
  {volume} {110}},\ \bibinfo {pages} {114} (\bibinfo {year}
  {2010})}\BibitemShut {NoStop}%
\bibitem [{\citenamefont {Marks}\ \emph {et~al.}(2014)\citenamefont {Marks},
  \citenamefont {Sch\"{o}ll},\ and\ \citenamefont
  {H\"{o}fer}}]{marks_formation_2014}%
  \BibitemOpen
  \bibfield  {author} {\bibinfo {author} {\bibfnamefont {M.}~\bibnamefont
  {Marks}}, \bibinfo {author} {\bibfnamefont {A.}~\bibnamefont {Sch\"{o}ll}}, \
  and\ \bibinfo {author} {\bibfnamefont {U.}~\bibnamefont {H\"{o}fer}},\
  }\href@noop {} {\bibfield  {journal} {\bibinfo  {journal} {Journal of
  Electron Spectroscopy and Related Phenomena}\ }\textbf {\bibinfo {volume}
  {195}},\ \bibinfo {pages} {263} (\bibinfo {year} {2014})}\BibitemShut
  {NoStop}%
\bibitem [{\citenamefont {Ferretti}\ \emph {et~al.}(2007)\citenamefont
  {Ferretti}, \citenamefont {Baldacchini}, \citenamefont {Calzolari},
  \citenamefont {Di~Felice}, \citenamefont {Ruini}, \citenamefont {Molinari},\
  and\ \citenamefont {Betti}}]{ferretti_mixing_2007}%
  \BibitemOpen
  \bibfield  {author} {\bibinfo {author} {\bibfnamefont {A.}~\bibnamefont
  {Ferretti}}, \bibinfo {author} {\bibfnamefont {C.}~\bibnamefont
  {Baldacchini}}, \bibinfo {author} {\bibfnamefont {A.}~\bibnamefont
  {Calzolari}}, \bibinfo {author} {\bibfnamefont {R.}~\bibnamefont
  {Di~Felice}}, \bibinfo {author} {\bibfnamefont {A.}~\bibnamefont {Ruini}},
  \bibinfo {author} {\bibfnamefont {E.}~\bibnamefont {Molinari}}, \ and\
  \bibinfo {author} {\bibfnamefont {M.~G.}\ \bibnamefont {Betti}},\ }\href@noop
  {} {\bibfield  {journal} {\bibinfo  {journal} {Physical Review Letters}\
  }\textbf {\bibinfo {volume} {99}},\ \bibinfo {pages} {046802} (\bibinfo
  {year} {2007})}\BibitemShut {NoStop}%
\bibitem [{\citenamefont {Gonzalez-Lakunza}\ \emph {et~al.}(2008)\citenamefont
  {Gonzalez-Lakunza}, \citenamefont {Fern\'{a}ndez-Torrente}, \citenamefont
  {Franke}, \citenamefont {Lorente}, \citenamefont {Arnau},\ and\ \citenamefont
  {Pascual}}]{gonzalez-lakunza_formation_2008}%
  \BibitemOpen
  \bibfield  {author} {\bibinfo {author} {\bibfnamefont {N.}~\bibnamefont
  {Gonzalez-Lakunza}}, \bibinfo {author} {\bibfnamefont {I.}~\bibnamefont
  {Fern\'{a}ndez-Torrente}}, \bibinfo {author} {\bibfnamefont {K.~J.}\
  \bibnamefont {Franke}}, \bibinfo {author} {\bibfnamefont {N.}~\bibnamefont
  {Lorente}}, \bibinfo {author} {\bibfnamefont {A.}~\bibnamefont {Arnau}}, \
  and\ \bibinfo {author} {\bibfnamefont {J.~I.}\ \bibnamefont {Pascual}},\
  }\href@noop {} {\bibfield  {journal} {\bibinfo  {journal} {Physical Review
  Letters}\ }\textbf {\bibinfo {volume} {100}},\ \bibinfo {pages} {156805}
  (\bibinfo {year} {2008})}\BibitemShut {NoStop}%
\bibitem [{\citenamefont {Ziroff}\ \emph {et~al.}(2010)\citenamefont {Ziroff},
  \citenamefont {Forster}, \citenamefont {Sch\"{o}ll}, \citenamefont
  {Puschnig},\ and\ \citenamefont {Reinert}}]{ziroff_hybridization_2010}%
  \BibitemOpen
  \bibfield  {author} {\bibinfo {author} {\bibfnamefont {J.}~\bibnamefont
  {Ziroff}}, \bibinfo {author} {\bibfnamefont {F.}~\bibnamefont {Forster}},
  \bibinfo {author} {\bibfnamefont {A.}~\bibnamefont {Sch\"{o}ll}}, \bibinfo
  {author} {\bibfnamefont {P.}~\bibnamefont {Puschnig}}, \ and\ \bibinfo
  {author} {\bibfnamefont {F.}~\bibnamefont {Reinert}},\ }\href@noop {}
  {\bibfield  {journal} {\bibinfo  {journal} {Physical Review Letters}\
  }\textbf {\bibinfo {volume} {104}},\ \bibinfo {pages} {233004} (\bibinfo
  {year} {2010})}\BibitemShut {NoStop}%
\bibitem [{\citenamefont {Marks}\ \emph {et~al.}(2011)\citenamefont {Marks},
  \citenamefont {Zaitsev}, \citenamefont {Schmidt}, \citenamefont {Schwalb},
  \citenamefont {Sch\"{o}ll}, \citenamefont {Nechaev}, \citenamefont
  {Echenique}, \citenamefont {Chulkov},\ and\ \citenamefont
  {H\"{o}fer}}]{marks_energy_2011}%
  \BibitemOpen
  \bibfield  {author} {\bibinfo {author} {\bibfnamefont {M.}~\bibnamefont
  {Marks}}, \bibinfo {author} {\bibfnamefont {N.~L.}\ \bibnamefont {Zaitsev}},
  \bibinfo {author} {\bibfnamefont {B.}~\bibnamefont {Schmidt}}, \bibinfo
  {author} {\bibfnamefont {C.~H.}\ \bibnamefont {Schwalb}}, \bibinfo {author}
  {\bibfnamefont {A.}~\bibnamefont {Sch\"{o}ll}}, \bibinfo {author}
  {\bibfnamefont {I.~A.}\ \bibnamefont {Nechaev}}, \bibinfo {author}
  {\bibfnamefont {P.~M.}\ \bibnamefont {Echenique}}, \bibinfo {author}
  {\bibfnamefont {E.~V.}\ \bibnamefont {Chulkov}}, \ and\ \bibinfo {author}
  {\bibfnamefont {U.}~\bibnamefont {H\"{o}fer}},\ }\href@noop {} {\bibfield
  {journal} {\bibinfo  {journal} {Phys. Rev. B}\ }\textbf {\bibinfo {volume}
  {84}},\ \bibinfo {pages} {081301} (\bibinfo {year} {2011})}\BibitemShut
  {NoStop}%
\bibitem [{\citenamefont {Galbraith}\ \emph {et~al.}(2014)\citenamefont
  {Galbraith}, \citenamefont {Marks}, \citenamefont {Tonner},\ and\
  \citenamefont {H\"{o}fer}}]{galbraith_formation_2014}%
  \BibitemOpen
  \bibfield  {author} {\bibinfo {author} {\bibfnamefont {M.~C.~E.}\
  \bibnamefont {Galbraith}}, \bibinfo {author} {\bibfnamefont {M.}~\bibnamefont
  {Marks}}, \bibinfo {author} {\bibfnamefont {R.}~\bibnamefont {Tonner}}, \
  and\ \bibinfo {author} {\bibfnamefont {U.}~\bibnamefont {H\"{o}fer}},\
  }\href@noop {} {\bibfield  {journal} {\bibinfo  {journal} {The Journal of
  Physical Chemistry Letters}\ }\textbf {\bibinfo {volume} {5}},\ \bibinfo
  {pages} {50} (\bibinfo {year} {2014})}\BibitemShut {NoStop}%
\bibitem [{\citenamefont {Caplins}\ \emph {et~al.}(2014)\citenamefont
  {Caplins}, \citenamefont {Suich}, \citenamefont {Shearer},\ and\
  \citenamefont {Harris}}]{caplins_metal/phthalocyanine_2014}%
  \BibitemOpen
  \bibfield  {author} {\bibinfo {author} {\bibfnamefont {B.~W.}\ \bibnamefont
  {Caplins}}, \bibinfo {author} {\bibfnamefont {D.~E.}\ \bibnamefont {Suich}},
  \bibinfo {author} {\bibfnamefont {A.~J.}\ \bibnamefont {Shearer}}, \ and\
  \bibinfo {author} {\bibfnamefont {C.~B.}\ \bibnamefont {Harris}},\
  }\href@noop {} {\bibfield  {journal} {\bibinfo  {journal} {The Journal of
  Physical Chemistry Letters}\ ,\ \bibinfo {pages} {1679}} (\bibinfo {year}
  {2014})}\BibitemShut {NoStop}%
\bibitem [{\citenamefont {Sch\"afer}\ \emph {et~al.}(2000)\citenamefont
  {Sch\"afer}, \citenamefont {Shumay}, \citenamefont {Wiets}, \citenamefont
  {Weinelt}, \citenamefont {Fauster}, \citenamefont {Chulkov}, \citenamefont
  {Silkin},\ and\ \citenamefont {Echenique}}]{schafer_lifetimes_2000}%
  \BibitemOpen
  \bibfield  {author} {\bibinfo {author} {\bibfnamefont {A.}~\bibnamefont
  {Sch\"afer}}, \bibinfo {author} {\bibfnamefont {I.~L.}\ \bibnamefont
  {Shumay}}, \bibinfo {author} {\bibfnamefont {M.}~\bibnamefont {Wiets}},
  \bibinfo {author} {\bibfnamefont {M.}~\bibnamefont {Weinelt}}, \bibinfo
  {author} {\bibfnamefont {T.}~\bibnamefont {Fauster}}, \bibinfo {author}
  {\bibfnamefont {E.~V.}\ \bibnamefont {Chulkov}}, \bibinfo {author}
  {\bibfnamefont {V.~M.}\ \bibnamefont {Silkin}}, \ and\ \bibinfo {author}
  {\bibfnamefont {P.~M.}\ \bibnamefont {Echenique}},\ }\href@noop {} {\bibfield
   {journal} {\bibinfo  {journal} {Phys. Rev. B}\ }\textbf {\bibinfo {volume}
  {61}},\ \bibinfo {pages} {13159} (\bibinfo {year} {2000})}\BibitemShut
  {NoStop}%
\bibitem [{\citenamefont {Dyer}\ and\ \citenamefont
  {Persson}(2010)}]{dyer_nature_2010}%
  \BibitemOpen
  \bibfield  {author} {\bibinfo {author} {\bibfnamefont {M.~S.}\ \bibnamefont
  {Dyer}}\ and\ \bibinfo {author} {\bibfnamefont {M.}~\bibnamefont {Persson}},\
  }\href@noop {} {\bibfield  {journal} {\bibinfo  {journal} {New Journal of
  Physics}\ }\textbf {\bibinfo {volume} {12}},\ \bibinfo {pages} {063014}
  (\bibinfo {year} {2010})}\BibitemShut {NoStop}%
\bibitem [{\citenamefont {Zaitsev}\ \emph {et~al.}(2012)\citenamefont
  {Zaitsev}, \citenamefont {Nechaev}, \citenamefont {Echenique},\ and\
  \citenamefont {Chulkov}}]{zaitsev_transformation_2012}%
  \BibitemOpen
  \bibfield  {author} {\bibinfo {author} {\bibfnamefont {N.~L.}\ \bibnamefont
  {Zaitsev}}, \bibinfo {author} {\bibfnamefont {I.~A.}\ \bibnamefont
  {Nechaev}}, \bibinfo {author} {\bibfnamefont {P.~M.}\ \bibnamefont
  {Echenique}}, \ and\ \bibinfo {author} {\bibfnamefont {E.~V.}\ \bibnamefont
  {Chulkov}},\ }\href@noop {} {\bibfield  {journal} {\bibinfo  {journal} {Phys.
  Rev. B}\ }\textbf {\bibinfo {volume} {85}},\ \bibinfo {pages} {115301}
  (\bibinfo {year} {2012})}\BibitemShut {NoStop}%
\bibitem [{\citenamefont {Chulkov}\ \emph {et~al.}(2001)\citenamefont
  {Chulkov}, \citenamefont {Silkin},\ and\ \citenamefont
  {Machado}}]{chulkov_quasiparticle_2001}%
  \BibitemOpen
  \bibfield  {author} {\bibinfo {author} {\bibfnamefont {E.~V.}\ \bibnamefont
  {Chulkov}}, \bibinfo {author} {\bibfnamefont {V.~M.}\ \bibnamefont {Silkin}},
  \ and\ \bibinfo {author} {\bibfnamefont {M.}~\bibnamefont {Machado}},\
  }\href@noop {} {\bibfield  {journal} {\bibinfo  {journal} {Surface Science}\
  }\textbf {\bibinfo {volume} {482--485}},\ \bibinfo {pages} {693} (\bibinfo
  {year} {2001})}\BibitemShut {NoStop}%
\bibitem [{\citenamefont {Willenbockel}\ \emph {et~al.}(2015)\citenamefont
  {Willenbockel}, \citenamefont {L\"{u}ftner}, \citenamefont {Stadtm\"{u}ller},
  \citenamefont {Koller}, \citenamefont {Kumpf}, \citenamefont {Soubatch},
  \citenamefont {Puschnig}, \citenamefont {Ramsey},\ and\ \citenamefont
  {Tautz}}]{willenbockel_interplay_2015}%
  \BibitemOpen
  \bibfield  {author} {\bibinfo {author} {\bibfnamefont {M.}~\bibnamefont
  {Willenbockel}}, \bibinfo {author} {\bibfnamefont {D.}~\bibnamefont
  {L\"{u}ftner}}, \bibinfo {author} {\bibfnamefont {B.}~\bibnamefont
  {Stadtm\"{u}ller}}, \bibinfo {author} {\bibfnamefont {G.}~\bibnamefont
  {Koller}}, \bibinfo {author} {\bibfnamefont {C.}~\bibnamefont {Kumpf}},
  \bibinfo {author} {\bibfnamefont {S.}~\bibnamefont {Soubatch}}, \bibinfo
  {author} {\bibfnamefont {P.}~\bibnamefont {Puschnig}}, \bibinfo {author}
  {\bibfnamefont {M.~G.}\ \bibnamefont {Ramsey}}, \ and\ \bibinfo {author}
  {\bibfnamefont {F.~S.}\ \bibnamefont {Tautz}},\ }\href@noop {} {\bibfield
  {journal} {\bibinfo  {journal} {Physical Chemistry Chemical Physics}\
  }\textbf {\bibinfo {volume} {17}},\ \bibinfo {pages} {1530} (\bibinfo {year}
  {2015})}\BibitemShut {NoStop}%
\bibitem [{\citenamefont {Boer}\ \emph {et~al.}(2003)\citenamefont {Boer},
  \citenamefont {Klapwijk},\ and\ \citenamefont
  {Morpurgo}}]{boer_field-effect_2003}%
  \BibitemOpen
  \bibfield  {author} {\bibinfo {author} {\bibfnamefont {R.~W. I.~d.}\
  \bibnamefont {Boer}}, \bibinfo {author} {\bibfnamefont {T.~M.}\ \bibnamefont
  {Klapwijk}}, \ and\ \bibinfo {author} {\bibfnamefont {A.~F.}\ \bibnamefont
  {Morpurgo}},\ }\href@noop {} {\bibfield  {journal} {\bibinfo  {journal}
  {Applied Physics Letters}\ }\textbf {\bibinfo {volume} {83}},\ \bibinfo
  {pages} {4345} (\bibinfo {year} {2003})}\BibitemShut {NoStop}%
\bibitem [{\citenamefont {Newman}\ \emph {et~al.}(2004)\citenamefont {Newman},
  \citenamefont {Chesterfield}, \citenamefont {Merlo},\ and\ \citenamefont
  {Frisbie}}]{newman_transport_2004}%
  \BibitemOpen
  \bibfield  {author} {\bibinfo {author} {\bibfnamefont {C.~R.}\ \bibnamefont
  {Newman}}, \bibinfo {author} {\bibfnamefont {R.~J.}\ \bibnamefont
  {Chesterfield}}, \bibinfo {author} {\bibfnamefont {J.~A.}\ \bibnamefont
  {Merlo}}, \ and\ \bibinfo {author} {\bibfnamefont {C.~D.}\ \bibnamefont
  {Frisbie}},\ }\href@noop {} {\bibfield  {journal} {\bibinfo  {journal}
  {Applied Physics Letters}\ }\textbf {\bibinfo {volume} {85}},\ \bibinfo
  {pages} {422} (\bibinfo {year} {2004})}\BibitemShut {NoStop}%
\bibitem [{\citenamefont {Soubatch}\ \emph {et~al.}(2011)\citenamefont
  {Soubatch}, \citenamefont {Kr\"{o}ger}, \citenamefont {Kumpf},\ and\
  \citenamefont {Tautz}}]{soubatch_structure_2011}%
  \BibitemOpen
  \bibfield  {author} {\bibinfo {author} {\bibfnamefont {S.}~\bibnamefont
  {Soubatch}}, \bibinfo {author} {\bibfnamefont {I.}~\bibnamefont
  {Kr\"{o}ger}}, \bibinfo {author} {\bibfnamefont {C.}~\bibnamefont {Kumpf}}, \
  and\ \bibinfo {author} {\bibfnamefont {F.}~\bibnamefont {Tautz}},\
  }\href@noop {} {\bibfield  {journal} {\bibinfo  {journal} {Phys. Rev. B}\
  }\textbf {\bibinfo {volume} {84}},\ \bibinfo {pages} {195440} (\bibinfo
  {year} {2011})}\BibitemShut {NoStop}%
\bibitem [{\citenamefont {Sueyoshi}\ \emph {et~al.}(2013)\citenamefont
  {Sueyoshi}, \citenamefont {Willenbockel}, \citenamefont {Naboka},
  \citenamefont {Nefedov}, \citenamefont {Soubatch}, \citenamefont {W\"{o}ll},\
  and\ \citenamefont {Tautz}}]{sueyoshi_spontaneous_2013}%
  \BibitemOpen
  \bibfield  {author} {\bibinfo {author} {\bibfnamefont {T.}~\bibnamefont
  {Sueyoshi}}, \bibinfo {author} {\bibfnamefont {M.}~\bibnamefont
  {Willenbockel}}, \bibinfo {author} {\bibfnamefont {M.}~\bibnamefont
  {Naboka}}, \bibinfo {author} {\bibfnamefont {A.}~\bibnamefont {Nefedov}},
  \bibinfo {author} {\bibfnamefont {S.}~\bibnamefont {Soubatch}}, \bibinfo
  {author} {\bibfnamefont {C.}~\bibnamefont {W\"{o}ll}}, \ and\ \bibinfo
  {author} {\bibfnamefont {F.~S.}\ \bibnamefont {Tautz}},\ }\href@noop {}
  {\bibfield  {journal} {\bibinfo  {journal} {The Journal of Physical Chemistry
  C}\ }\textbf {\bibinfo {volume} {117}},\ \bibinfo {pages} {9212} (\bibinfo
  {year} {2013})}\BibitemShut {NoStop}%
\bibitem [{\citenamefont {Langner}\ \emph {et~al.}(2005)\citenamefont
  {Langner}, \citenamefont {Hauschild}, \citenamefont {Fahrenholz},\ and\
  \citenamefont {Sokolowski}}]{langner_structural_2005}%
  \BibitemOpen
  \bibfield  {author} {\bibinfo {author} {\bibfnamefont {A.}~\bibnamefont
  {Langner}}, \bibinfo {author} {\bibfnamefont {A.}~\bibnamefont {Hauschild}},
  \bibinfo {author} {\bibfnamefont {S.}~\bibnamefont {Fahrenholz}}, \ and\
  \bibinfo {author} {\bibfnamefont {M.}~\bibnamefont {Sokolowski}},\
  }\href@noop {} {\bibfield  {journal} {\bibinfo  {journal} {Surface science}\
  }\textbf {\bibinfo {volume} {574}},\ \bibinfo {pages} {153} (\bibinfo {year}
  {2005})}\BibitemShut {NoStop}%
\bibitem [{\citenamefont {Gonella}\ \emph {et~al.}(2008)\citenamefont
  {Gonella}, \citenamefont {Dai},\ and\ \citenamefont
  {Rockey}}]{gonella_tetracene_2008}%
  \BibitemOpen
  \bibfield  {author} {\bibinfo {author} {\bibfnamefont {G.}~\bibnamefont
  {Gonella}}, \bibinfo {author} {\bibfnamefont {H.-L.}\ \bibnamefont {Dai}}, \
  and\ \bibinfo {author} {\bibfnamefont {T.}~\bibnamefont {Rockey}},\
  }\href@noop {} {\bibfield  {journal} {\bibinfo  {journal} {Journal of
  Physical Chemistry C}\ }\textbf {\bibinfo {volume} {112}},\ \bibinfo {pages}
  {4696} (\bibinfo {year} {2008})}\BibitemShut {NoStop}%
\bibitem [{\citenamefont {Garc\'{i}a-Gil}\ \emph {et~al.}(2009)\citenamefont
  {Garc\'{i}a-Gil}, \citenamefont {Garc\'{i}a}, \citenamefont {Lorente},\ and\
  \citenamefont {Ordej\'{o}n}}]{garcia-gil_optimal_2009}%
  \BibitemOpen
  \bibfield  {author} {\bibinfo {author} {\bibfnamefont {S.}~\bibnamefont
  {Garc\'{i}a-Gil}}, \bibinfo {author} {\bibfnamefont {A.}~\bibnamefont
  {Garc\'{i}a}}, \bibinfo {author} {\bibfnamefont {N.}~\bibnamefont {Lorente}},
  \ and\ \bibinfo {author} {\bibfnamefont {P.}~\bibnamefont {Ordej\'{o}n}},\
  }\href@noop {} {\bibfield  {journal} {\bibinfo  {journal} {Phys. Rev. B}\
  }\textbf {\bibinfo {volume} {79}},\ \bibinfo {pages} {075441} (\bibinfo
  {year} {2009})}\BibitemShut {NoStop}%
\bibitem [{\citenamefont {Ordej\'{o}n}\ \emph {et~al.}(1996)\citenamefont
  {Ordej\'{o}n}, \citenamefont {Artacho},\ and\ \citenamefont
  {Soler}}]{ordejon_self-consistent_1996}%
  \BibitemOpen
  \bibfield  {author} {\bibinfo {author} {\bibfnamefont {P.}~\bibnamefont
  {Ordej\'{o}n}}, \bibinfo {author} {\bibfnamefont {E.}~\bibnamefont
  {Artacho}}, \ and\ \bibinfo {author} {\bibfnamefont {J.~M.}\ \bibnamefont
  {Soler}},\ }\href@noop {} {\bibfield  {journal} {\bibinfo  {journal}
  {Physical Review B}\ }\textbf {\bibinfo {volume} {53}},\ \bibinfo {pages}
  {R10441} (\bibinfo {year} {1996})}\BibitemShut {NoStop}%
\bibitem [{\citenamefont {Soler}\ \emph {et~al.}(2002)\citenamefont {Soler},
  \citenamefont {Artacho}, \citenamefont {Gale}, \citenamefont {Garc\'{i}a},
  \citenamefont {Junquera}, \citenamefont {Ordej\'{o}n},\ and\ \citenamefont
  {{S\'{a}nchez-Portal}}}]{soler_siesta_2002}%
  \BibitemOpen
  \bibfield  {author} {\bibinfo {author} {\bibfnamefont {J.~M.}\ \bibnamefont
  {Soler}}, \bibinfo {author} {\bibfnamefont {E.}~\bibnamefont {Artacho}},
  \bibinfo {author} {\bibfnamefont {J.~D.}\ \bibnamefont {Gale}}, \bibinfo
  {author} {\bibfnamefont {A.}~\bibnamefont {Garc\'{i}a}}, \bibinfo {author}
  {\bibfnamefont {J.}~\bibnamefont {Junquera}}, \bibinfo {author}
  {\bibfnamefont {P.}~\bibnamefont {Ordej\'{o}n}}, \ and\ \bibinfo {author}
  {\bibfnamefont {D.}~\bibnamefont {{S\'{a}nchez-Portal}}},\ }\href@noop {}
  {\bibfield  {journal} {\bibinfo  {journal} {Journal of Physics: Condensed
  Matter}\ }\textbf {\bibinfo {volume} {14}},\ \bibinfo {pages} {2745}
  (\bibinfo {year} {2002})}\BibitemShut {NoStop}%
\bibitem [{\citenamefont {Troullier}\ and\ \citenamefont
  {Martins}(1991)}]{troullier_efficient_1991}%
  \BibitemOpen
  \bibfield  {author} {\bibinfo {author} {\bibfnamefont {N.}~\bibnamefont
  {Troullier}}\ and\ \bibinfo {author} {\bibfnamefont {J.~L.}\ \bibnamefont
  {Martins}},\ }\href@noop {} {\bibfield  {journal} {\bibinfo  {journal}
  {Physical Review B}\ }\textbf {\bibinfo {volume} {43}},\ \bibinfo {pages}
  {1993} (\bibinfo {year} {1991})}\BibitemShut {NoStop}%
\bibitem [{\citenamefont {Kleinman}\ and\ \citenamefont
  {Bylander}(1982)}]{kleinman_pseudopotentials_1982}%
  \BibitemOpen
  \bibfield  {author} {\bibinfo {author} {\bibfnamefont {L.}~\bibnamefont
  {Kleinman}}\ and\ \bibinfo {author} {\bibfnamefont {D.~M.}\ \bibnamefont
  {Bylander}},\ }\href@noop {} {\bibfield  {journal} {\bibinfo  {journal}
  {Phys. Rev. Lett.}\ }\textbf {\bibinfo {volume} {48}},\ \bibinfo {pages}
  {1425} (\bibinfo {year} {1982})}\BibitemShut {NoStop}%
\bibitem [{\citenamefont {Perdew}\ \emph {et~al.}(1996)\citenamefont {Perdew},
  \citenamefont {Burke},\ and\ \citenamefont
  {Ernzerhof}}]{perdew_generalized-gradient_1996}%
  \BibitemOpen
  \bibfield  {author} {\bibinfo {author} {\bibfnamefont {J.~P.}\ \bibnamefont
  {Perdew}}, \bibinfo {author} {\bibfnamefont {K.}~\bibnamefont {Burke}}, \
  and\ \bibinfo {author} {\bibfnamefont {M.}~\bibnamefont {Ernzerhof}},\
  }\href@noop {} {\bibfield  {journal} {\bibinfo  {journal} {Phys. Rev. Lett.}\
  }\textbf {\bibinfo {volume} {77}},\ \bibinfo {pages} {3865} (\bibinfo {year}
  {1996})}\BibitemShut {NoStop}%
\bibitem [{\citenamefont {Klime\v{s}}\ \emph {et~al.}(2010)\citenamefont
  {Klime\v{s}}, \citenamefont {Bowler},\ and\ \citenamefont
  {Michaelides}}]{klimes_chemical_2010}%
  \BibitemOpen
  \bibfield  {author} {\bibinfo {author} {\bibfnamefont {J.}~\bibnamefont
  {Klime\v{s}}}, \bibinfo {author} {\bibfnamefont {D.~R.}\ \bibnamefont
  {Bowler}}, \ and\ \bibinfo {author} {\bibfnamefont {A.}~\bibnamefont
  {Michaelides}},\ }\href@noop {} {\bibfield  {journal} {\bibinfo  {journal}
  {Journal of Physics: Condensed Matter}\ }\textbf {\bibinfo {volume} {22}},\
  \bibinfo {pages} {022201} (\bibinfo {year} {2010})}\BibitemShut {NoStop}%
\bibitem [{\citenamefont {Klime\v{s}}\ \emph {et~al.}(2011)\citenamefont
  {Klime\v{s}}, \citenamefont {Bowler},\ and\ \citenamefont
  {Michaelides}}]{klimes_van_2011}%
  \BibitemOpen
  \bibfield  {author} {\bibinfo {author} {\bibfnamefont {J.}~\bibnamefont
  {Klime\v{s}}}, \bibinfo {author} {\bibfnamefont {D.~R.}\ \bibnamefont
  {Bowler}}, \ and\ \bibinfo {author} {\bibfnamefont {A.}~\bibnamefont
  {Michaelides}},\ }\href@noop {} {\bibfield  {journal} {\bibinfo  {journal}
  {Physical Review B}\ }\textbf {\bibinfo {volume} {83}},\ \bibinfo {pages}
  {195131} (\bibinfo {year} {2011})}\BibitemShut {NoStop}%
\bibitem [{\citenamefont {Lee}\ and\ \citenamefont {Yu}(2005)}]{lee_ab_2005}%
  \BibitemOpen
  \bibfield  {author} {\bibinfo {author} {\bibfnamefont {K.}~\bibnamefont
  {Lee}}\ and\ \bibinfo {author} {\bibfnamefont {J.}~\bibnamefont {Yu}},\
  }\href@noop {} {\bibfield  {journal} {\bibinfo  {journal} {Surface Science}\
  }\textbf {\bibinfo {volume} {589}},\ \bibinfo {pages} {8} (\bibinfo {year}
  {2005})}\BibitemShut {NoStop}%
\bibitem [{\citenamefont {Buimaga-Iarinca}\ and\ \citenamefont
  {Morari}(2014)}]{buimaga-iarinca_adsorption_2014}%
  \BibitemOpen
  \bibfield  {author} {\bibinfo {author} {\bibfnamefont {L.}~\bibnamefont
  {Buimaga-Iarinca}}\ and\ \bibinfo {author} {\bibfnamefont {C.}~\bibnamefont
  {Morari}},\ }\href@noop {} {\bibfield  {journal} {\bibinfo  {journal}
  {Theoretical Chemistry Accounts}\ }\textbf {\bibinfo {volume} {133}}
  (\bibinfo {year} {2014})}\BibitemShut {NoStop}%
\bibitem [{\citenamefont {Johnson}(1988)}]{johnson_modified_1988}%
  \BibitemOpen
  \bibfield  {author} {\bibinfo {author} {\bibfnamefont {D.~D.}\ \bibnamefont
  {Johnson}},\ }\href@noop {} {\bibfield  {journal} {\bibinfo  {journal} {Phys.
  Rev. B}\ }\textbf {\bibinfo {volume} {38}},\ \bibinfo {pages} {12807}
  (\bibinfo {year} {1988})}\BibitemShut {NoStop}%
\bibitem [{\citenamefont {Ozaki}\ \emph {et~al.}()\citenamefont {Ozaki},
  \citenamefont {Kino}, \citenamefont {Yu}, \citenamefont {Han}, \citenamefont
  {Kobayashi}, \citenamefont {Ohfuti}, \citenamefont {Ishii}, \citenamefont
  {Ohwaki}, \citenamefont {Weng},\ and\ \citenamefont {Terakura}}]{_openmx_37}%
  \BibitemOpen
  \bibfield  {author} {\bibinfo {author} {\bibfnamefont {T.}~\bibnamefont
  {Ozaki}}, \bibinfo {author} {\bibfnamefont {H.}~\bibnamefont {Kino}},
  \bibinfo {author} {\bibfnamefont {J.}~\bibnamefont {Yu}}, \bibinfo {author}
  {\bibfnamefont {M.}~\bibnamefont {Han}}, \bibinfo {author} {\bibfnamefont
  {N.}~\bibnamefont {Kobayashi}}, \bibinfo {author} {\bibfnamefont
  {M.}~\bibnamefont {Ohfuti}}, \bibinfo {author} {\bibfnamefont
  {F.}~\bibnamefont {Ishii}}, \bibinfo {author} {\bibfnamefont
  {T.}~\bibnamefont {Ohwaki}}, \bibinfo {author} {\bibfnamefont
  {H.}~\bibnamefont {Weng}}, \ and\ \bibinfo {author} {\bibfnamefont
  {K.}~\bibnamefont {Terakura}},\ }\href@noop {} {}\bibinfo {howpublished}
  {\url{http://openmx-square.org/}}\BibitemShut {NoStop}%
\bibitem [{\citenamefont {Ozaki}(2003)}]{ozaki_variationally_2003}%
  \BibitemOpen
  \bibfield  {author} {\bibinfo {author} {\bibfnamefont {T.}~\bibnamefont
  {Ozaki}},\ }\href@noop {} {\bibfield  {journal} {\bibinfo  {journal}
  {Physical Review B}\ }\textbf {\bibinfo {volume} {67}},\ \bibinfo {pages}
  {155108} (\bibinfo {year} {2003})}\BibitemShut {NoStop}%
\bibitem [{\citenamefont {Ozaki}\ and\ \citenamefont
  {Kino}(2004)}]{ozaki_numerical_2004}%
  \BibitemOpen
  \bibfield  {author} {\bibinfo {author} {\bibfnamefont {T.}~\bibnamefont
  {Ozaki}}\ and\ \bibinfo {author} {\bibfnamefont {H.}~\bibnamefont {Kino}},\
  }\href@noop {} {\bibfield  {journal} {\bibinfo  {journal} {Physical Review
  B}\ }\textbf {\bibinfo {volume} {69}},\ \bibinfo {pages} {195113} (\bibinfo
  {year} {2004})}\BibitemShut {NoStop}%
\bibitem [{\citenamefont {Ozaki}\ and\ \citenamefont
  {Kino}(2005)}]{ozaki_efficient_2005}%
  \BibitemOpen
  \bibfield  {author} {\bibinfo {author} {\bibfnamefont {T.}~\bibnamefont
  {Ozaki}}\ and\ \bibinfo {author} {\bibfnamefont {H.}~\bibnamefont {Kino}},\
  }\href@noop {} {\bibfield  {journal} {\bibinfo  {journal} {Physical Review
  B}\ }\textbf {\bibinfo {volume} {72}},\ \bibinfo {pages} {045121} (\bibinfo
  {year} {2005})}\BibitemShut {NoStop}%
\bibitem [{\citenamefont {Boys}\ and\ \citenamefont
  {Bernardi}(1970)}]{boys_calculation_1970}%
  \BibitemOpen
  \bibfield  {author} {\bibinfo {author} {\bibfnamefont {S.}~\bibnamefont
  {Boys}}\ and\ \bibinfo {author} {\bibfnamefont {F.}~\bibnamefont
  {Bernardi}},\ }\href@noop {} {\bibfield  {journal} {\bibinfo  {journal}
  {Molecular Physics}\ }\textbf {\bibinfo {volume} {19}},\ \bibinfo {pages}
  {553} (\bibinfo {year} {1970})}\BibitemShut {NoStop}%
\bibitem [{\citenamefont {Tao}\ \emph {et~al.}(2015)\citenamefont {Tao},
  \citenamefont {Mao}, \citenamefont {Zhang},\ and\ \citenamefont
  {He}}]{tao_electronic_2015}%
  \BibitemOpen
  \bibfield  {author} {\bibinfo {author} {\bibfnamefont {Y.}~\bibnamefont
  {Tao}}, \bibinfo {author} {\bibfnamefont {H.}~\bibnamefont {Mao}}, \bibinfo
  {author} {\bibfnamefont {H.}~\bibnamefont {Zhang}}, \ and\ \bibinfo {author}
  {\bibfnamefont {P.}~\bibnamefont {He}},\ }\href@noop {} {\bibfield  {journal}
  {\bibinfo  {journal} {Surface Science}\ }\textbf {\bibinfo {volume} {641}},\
  \bibinfo {pages} {135} (\bibinfo {year} {2015})}\BibitemShut {NoStop}%
\bibitem [{\citenamefont {Witte}\ \emph {et~al.}(2005)\citenamefont {Witte},
  \citenamefont {Lukas}, \citenamefont {Bagus},\ and\ \citenamefont
  {W\"{o}ll}}]{witte_vacuum_2005}%
  \BibitemOpen
  \bibfield  {author} {\bibinfo {author} {\bibfnamefont {G.}~\bibnamefont
  {Witte}}, \bibinfo {author} {\bibfnamefont {S.}~\bibnamefont {Lukas}},
  \bibinfo {author} {\bibfnamefont {P.~S.}\ \bibnamefont {Bagus}}, \ and\
  \bibinfo {author} {\bibfnamefont {C.}~\bibnamefont {W\"{o}ll}},\ }\href@noop
  {} {\bibfield  {journal} {\bibinfo  {journal} {Applied Physics Letters}\
  }\textbf {\bibinfo {volume} {87}},\ \bibinfo {pages} {263502} (\bibinfo
  {year} {2005})}\BibitemShut {NoStop}%
\bibitem [{\citenamefont {Romaner}\ \emph {et~al.}(2009)\citenamefont
  {Romaner}, \citenamefont {Nabok}, \citenamefont {Puschnig}, \citenamefont
  {Zojer},\ and\ \citenamefont {Ambrosch-Draxl}}]{romaner_theoretical_2009}%
  \BibitemOpen
  \bibfield  {author} {\bibinfo {author} {\bibfnamefont {L.}~\bibnamefont
  {Romaner}}, \bibinfo {author} {\bibfnamefont {D.}~\bibnamefont {Nabok}},
  \bibinfo {author} {\bibfnamefont {P.}~\bibnamefont {Puschnig}}, \bibinfo
  {author} {\bibfnamefont {E.}~\bibnamefont {Zojer}}, \ and\ \bibinfo {author}
  {\bibfnamefont {C.}~\bibnamefont {Ambrosch-Draxl}},\ }\href@noop {}
  {\bibfield  {journal} {\bibinfo  {journal} {New Journal of Physics}\ }\textbf
  {\bibinfo {volume} {11}},\ \bibinfo {pages} {053010} (\bibinfo {year}
  {2009})}\BibitemShut {NoStop}%
\bibitem [{\citenamefont {Rusu}\ \emph {et~al.}(2010)\citenamefont {Rusu},
  \citenamefont {Giovannetti}, \citenamefont {Weijtens}, \citenamefont
  {Coehoorn},\ and\ \citenamefont {Brocks}}]{rusu_first-principles_2010}%
  \BibitemOpen
  \bibfield  {author} {\bibinfo {author} {\bibfnamefont {P.~C.}\ \bibnamefont
  {Rusu}}, \bibinfo {author} {\bibfnamefont {G.}~\bibnamefont {Giovannetti}},
  \bibinfo {author} {\bibfnamefont {C.}~\bibnamefont {Weijtens}}, \bibinfo
  {author} {\bibfnamefont {R.}~\bibnamefont {Coehoorn}}, \ and\ \bibinfo
  {author} {\bibfnamefont {G.}~\bibnamefont {Brocks}},\ }\href@noop {}
  {\bibfield  {journal} {\bibinfo  {journal} {Phys. Rev. B}\ }\textbf {\bibinfo
  {volume} {81}},\ \bibinfo {pages} {125403} (\bibinfo {year}
  {2010})}\BibitemShut {NoStop}%
\bibitem [{\citenamefont {Frank}\ \emph {et~al.}(1988)\citenamefont {Frank},
  \citenamefont {Yannoulis}, \citenamefont {Dudde},\ and\ \citenamefont
  {Koch}}]{frank_unoccupied_1988}%
  \BibitemOpen
  \bibfield  {author} {\bibinfo {author} {\bibfnamefont {K.~H.}\ \bibnamefont
  {Frank}}, \bibinfo {author} {\bibfnamefont {P.}~\bibnamefont {Yannoulis}},
  \bibinfo {author} {\bibfnamefont {R.}~\bibnamefont {Dudde}}, \ and\ \bibinfo
  {author} {\bibfnamefont {E.~E.}\ \bibnamefont {Koch}},\ }\href@noop {}
  {\bibfield  {journal} {\bibinfo  {journal} {The Journal of Chemical Physics}\
  }\textbf {\bibinfo {volume} {89}},\ \bibinfo {pages} {7569} (\bibinfo {year}
  {1988})}\BibitemShut {NoStop}%
\bibitem [{\citenamefont {Scheybal}\ \emph {et~al.}(2009)\citenamefont
  {Scheybal}, \citenamefont {M\"{u}ller}, \citenamefont {Bertschinger},
  \citenamefont {Wahl}, \citenamefont {Bendounan}, \citenamefont {Aebi},\ and\
  \citenamefont {Jung}}]{scheybal_modification_2009}%
  \BibitemOpen
  \bibfield  {author} {\bibinfo {author} {\bibfnamefont {A.}~\bibnamefont
  {Scheybal}}, \bibinfo {author} {\bibfnamefont {K.}~\bibnamefont
  {M\"{u}ller}}, \bibinfo {author} {\bibfnamefont {R.}~\bibnamefont
  {Bertschinger}}, \bibinfo {author} {\bibfnamefont {M.}~\bibnamefont {Wahl}},
  \bibinfo {author} {\bibfnamefont {A.}~\bibnamefont {Bendounan}}, \bibinfo
  {author} {\bibfnamefont {P.}~\bibnamefont {Aebi}}, \ and\ \bibinfo {author}
  {\bibfnamefont {T.~A.}\ \bibnamefont {Jung}},\ }\href@noop {} {\bibfield
  {journal} {\bibinfo  {journal} {Physical Review B}\ }\textbf {\bibinfo
  {volume} {79}},\ \bibinfo {pages} {115406} (\bibinfo {year}
  {2009})}\BibitemShut {NoStop}%
\end{thebibliography}%

\end{document}